\documentclass[%
 aip,
 amsmath,amssymb,
 reprint,%
]{revtex4-1}

\usepackage{graphicx}
\usepackage{dcolumn}
\usepackage{bm}
\usepackage{bbm}
\usepackage{float}
\makeatletter
\let\newfloat\newfloat@ltx
\makeatother
\usepackage{algorithm}
\usepackage{algorithmic}
\usepackage[dvipsnames]{xcolor}

\usepackage{graphicx}
\usepackage{dcolumn}
\usepackage{bm}
\usepackage{dsfont}
\usepackage[hidelinks]{hyperref}
\usepackage[capitalise]{cleveref}
\usepackage{color}

\definecolor{myred}{cmyk}{0, 0.47, 0.38, 0.53}
\definecolor{myyellow}{cmyk}{0, 0.26, 0.75, 0.03}
\definecolor{mydarkorange}{cmyk}{0, 0.37, 0.71, 0.14}
\definecolor{mydarkorange}{cmyk}{0, 0.51, 0.68, 0.15}
\makeatletter
\def\@email#1#2{%
 \endgroup
 \patchcmd{\titleblock@produce}
  {\frontmatter@RRAPformat}
  {\frontmatter@RRAPformat{\produce@RRAP{*#1\href{mailto:#2}{#2}}}\frontmatter@RRAPformat}
  {}{}
}%
\makeatother
\begin{document}

\preprint{APS/123-QED}

\title{Revisiting Kinetic Monte Carlo Algorithms for Time-dependent Processes: from open-loop control to feedback control}

\author{Supraja S. Chittari}
\author{Zhiyue Lu}%
 \email{zhiyuelu@unc.edu}
\affiliation{%
Department of Chemistry, University of North Carolina-Chapel Hill, NC, USA}%

\date{\today}

\begin{abstract}
Simulating stochastic systems with feedback control is challenging due to the complex interplay between the system's dynamics and the feedback-dependent control protocols. We present a single-step-trajectory probability analysis to time-dependent stochastic systems. Based on this analysis, we revisit several time-dependent kinetic Monte Carlo (KMC) algorithms designed for systems under open-loop-control protocols. Our analysis provides an unified alternative proof to these algorithms, summarized into a pedagogical tutorial. Moreover, with the trajectory probability analysis, we present a novel feedback-controlled KMC algorithm that accurately captures the dynamics systems controlled by external signal based on measurements of the system's state. Our method correctly captures the system dynamics and avoids the artificial Zeno effect that arises from incorrectly applying the direct Gillespie algorithm to feedback-controlled systems. This work provides a unified perspective on existing open-loop-control KMC algorithms and also offers a powerful and accurate tool for simulating stochastic systems with feedback control.
\end{abstract}

\maketitle

\section{Introduction}
At the molecular scale, biological processes are complex and stochastic.\cite{shahrezaei2008stochastic, levine2007stochastic} Currently, numerical simulations contribute critical insights into the stochastic dynamics of the systems.\cite{szekely2014stochastic} One example approach involves solving for the master equation,\cite{earnshaw2010invariant, lee2010multi,meister2014modeling} constructed either through molecular dynamics simulations (e.g., via Markov state model approaches \cite{Anderson2011-gc, Bowman2013-ka, husic2018markov}) or experimental results (e.g., kinetic proofreading\cite{nelson2004biological,alon2019introduction}). In this case, if the system is reduced to a few state Markov network, one can numerically solve for the master equation, given the initial probability distribution and the control protocol. Another example, when the system's state space is too large, is to utilize the Gillespie Algorithm\cite{gillespie2001approximate} or kinetic Monte Carlo (KMC) simulation to obtain an ensemble of the stochastic trajectories of the process. 

Numerous biological processes also respond either to external control or other coupled processes.\cite{iglesias2010control} These stochastic systems are more complex than stationary systems, as their kinetic rates changes over time via open-loop (non-feedback) or closed-loop (feedback) control. Examples of feedback controlled systems span multiple length- and time-scales, from metabolic regulation\cite{lorendeau2015metabolic, atkinson1965biological, Goyal2010-ru} to bacterial chemotaxis\cite{yi2000robust, Hamadeh2011-hr, Yi2000-pk} to cell-cycle regulation.\cite{murray1992creative, domian1999feedback, cross1991potential, li1991feedback} Understanding feedback controlled systems is essential in unraveling their dynamics and optimizing their performance. However, it remains challenging to simulate feedback controlled stochastic systems due to the presence of in-time feedback.

Gillespie first proposed a direct KMC method that allows one to simulate the stochastic dynamics for a system that evolves under a constant set of rates.\cite{gillespie1977exact} Unfortunately, this method cannot be directly extended to simulate time-inhomogeneous systems, where the system's dynamical rates changes over time. Naively implementing the direct KMC simulation by discretizing the simulation into a sequence of constant rate windows may lead to an incorrect underestimation of the transition frequency of the systems, which resembles the Zeno effect. In quantum mechanics, the Zeno effect describes the system dynamics being trapped into a single state due to frequent repetitive projective measurements.\cite{Mobus2019-qx, itano1990quantum, streed2006continuous} 

In the past few decades, open-loop-control KMC algorithm has been proposed to simulate various time-inhomogeneous biological systems whose kinetic rates changes over time due to an open-loop (non-feedback) control protocol. These works were presented for various biological processes including ion channels in neurons,\cite{Anderson2015-ql, Ding2016-qr, Chow1996-ab} RNA folding,\cite{Li2009-ee} and chemical reactions.\cite{Anderson2007-dq, Anderson2011-gc} In 2007, Anderson proposed a modified next-reaction simulation method, where the transition rate of one possible reaction could change over time due to the occurrence of another possible reaction that takes place in a shorter time.\cite{Anderson2007-dq} In this case, the simulation addresses the changes of the reaction rates by providing a modification to the generation of the waiting time of reactions. Later, a general class of methods\cite{Anderson2007-dq,Anderson2015-ql, Anderson2011-gc, Chow1996-ab, Li2009-ee} were formulated for simulating systems with time-dependent rates. These works address systems that fall into the category of open-loop controlled stochastic systems,\cite{bechhoefer2021control} where the system's transition rates are controlled by external signal as a function of time according to a given protocol. However, still missing is a general theory and algorithm for simulating feedback controlled stochastic systems, where the protocol is updated at each scheduled measurement time based on the instantaneous measurement outcome.\cite{franklin2002feedback} The main difficulty in developing a KMC algorithm for feedback controlled systems is that the control protocol at future times is unknown, as it depends on the results of the future measurements of the system.  


In this work, we first provide an alternative derivation for the open-loop-controlled KMC, which recapitulates the methods developed in a number of existing references.\cite{Anderson2007-dq,Anderson2015-ql, Anderson2011-gc, Chow1996-ab, Li2009-ee} This alternative derivation, based on single-trajectory probability analysis, lays the foundation to the development of a novel algorithm able to simulate systems under in-time feedback control. The validity and the implementation of the algorithm is demonstrated by a Maxwell's feedback control refrigerator. 



\section{Theory and Algorithms}
\subsection{Theory: Single-step trajectory probability}\label{subsec: single-step}
This article begins by revisiting the theoretical foundation of the open-loop-control KMC method by providing a pedagogical proof that recapitulates various versions of these algorithms.\cite{Anderson2007-dq,Anderson2015-ql, Anderson2011-gc, Chow1996-ab, Li2009-ee} The theory is constructed by comparing the true probability of single-step trajectories and the simulated trajectory probability generated within one step of a modified KMC algorithm. This trajectory analysis not only provides an alternative proof for open-loop-control KMC methods described previously\cite{Anderson2007-dq,Anderson2015-ql, Anderson2011-gc, Chow1996-ab, Li2009-ee} but also provides the foundation from which a closed-loop-control KMC method can be derived. 


Consider a continuous-time discrete-state Markov process that is subject to an external control protocol $\lambda(t)$ as an arbitrary function of time $t$. The system contains a discrete number of states $n_s$ and is initialized with probability distribution $\vec p(t=0)$. The dynamics of $\vec p(t)$ is characterized by the master equation
\begin{equation}
    \frac{d \vec p}{dt} = \hat R(\lambda (t)) \vec p
\end{equation}
where matrix element $R_{ji}(\lambda (t))$ is the transient transition probability rates from $i$ to $j$ given the control parameter $\lambda$ at time $t$. The diagonal element satisfying $R_{ii}(t) = -\sum_{j \neq i} R_{ji}(t)$ is the negative transient escape rate from state $i$ at time $t$. The master equation evolution of $\vec p(t)$ can also be expressed in terms of the dynamics of $i$-th state's probability:
\begin{equation}
    \frac{{\rm d} p_i}{ {\rm d} t} = \sum_{j \neq i}R_{ij}(\lambda(t)) p_j(t)   + R_{ii}(\lambda(t)) p_i(t)
\end{equation}
For an open-loop controlled system, the control protocol $\lambda(t)$ is externally determined. In contrast, for a closed-loop controlled system (e.g., one with a series of feedback measurements), the specific form of $\lambda(t)$ may depend on the system's state. 

Each individual step of a KMC simulation is concerned with only generating the next stochastic event (i.e., evolving to state state $x_{k+1}$ at time $t_{k+1}$ from the previous state $x_k$ and time $t_{k}$). This event can be expressed as a single-step trajectory $X_{\rm single}=(x_{k},t_{k}, x', t')$ given the previous step. The conditional probability for this single-step trajectory can be written as
\begin{equation}
\label{eqn:trajectory}
    p[x',t' \vert x_{k},t_{k}] =  e^{\int_{t_k}^{t'} {R_{x_{k} x_{k}}(t) {\rm d}t}} R_{x' x_k}(t')
\end{equation}
which is also a joint probability distribution of the next event's time $t'$ and state $x'$. This formula is true for any Markov dynamics regardless of whether the dynamical rates are constants or the rates change over time. We show below that, in the case of time-independent or time-dependent rates, the direct Gillespie Algorithm or an open-loop-control KMC algorithm, respectively, can correctly capture the dynamics.

{\it Time-independent rates:} In the case where the dynamics of the system are constant over time \footnote{Here constant rate over time is used to describe the time period between two consecutive stochastic events. Naturally, when the system state changes after the next event, the possible transitions and their rates are naturally updated.}, the direct Gillespie Algorithm\cite{gillespie1977exact} is sufficient. From a probability theory perspective, the single-step transition probability of the next state $x'$ and next time $t'$ can be factored into the product of two statistically independent distributions and can be generated separately from their own probability distributions $P_x({x'})$ and $P_t({t'})$.
\begin{equation}
\label{eqn:trad_traj}
\begin{aligned}
    p[{x'},{t'} \vert x_{k},t_{k}] &= P_t({t'}) \cdot P_x({x'}) ~~, ~t' >t_k,~x' \neq x_k \\
    &=  e^{ ({t'}-t_k) \cdot {R_{x_{k} x_{k}} }} \cdot R_{{x'} x_k}
\end{aligned}
\end{equation}
In this case, the direct KMC \cite{gillespie1977exact} is carried out by generating the next transition time $t'$ according to $P_t({t'}) \propto e^{ {R_{x_{k} x_{k}} \cdot ({t'}-t_k)}}$ and generating the next state $x'$ according to the state probability $P_x({x'})\propto R_{{x'} x_k}$.

{\it Time-dependent rates:} If the rates change over time, the direct KMC approach\cite{gillespie1977exact} fails to generate the correct dynamics since the next event's time and state are statistically correlated. In this case, we can still factorize the joint probability of $t'$ and $x'$, $P({x'},{t'})\equiv p[{x'},{t'} \vert x_{k},t_{k}]$ as the product of two probability distributions:
\begin{equation}
    \label{eq:joint_x't'}
        P({x'},{t'})=P^{\text{marg}}_t(t')\cdot P^{\text{cond}}_x({x'}\vert {t'})
\end{equation}
where the marginal distribution of the transition time $t'$ (given the previous $x_k$ and $t_k$) is denoted by
\begin{equation}
    \label{eq:marg_t}
    \begin{aligned}
        P^{\text{marg}}_t(t') &= \sum_{{x'}\neq x_k} p[{x'},t' \vert x_{k},t_{k}]\\
        &=- e^{\int_{t_k}^{t'} R_{x_k x_k}(t) dt} \cdot R_{x_k x_k}(t')
    \end{aligned}
\end{equation}

and the conditional probability distribution of next state $x'$ conditioned on the transition time being $t'$ is 
\begin{equation}
    \label{eq:Pcond_x}
    P^{\text{cond}}_x({x'}\vert {t'}) = \frac{R_{{x'} x_k}(t')}{-R_{x_{k} x_k}(t')}
\end{equation}

In this case, a modified KMC algorithm should be capable of faithfully generating the stochastic trajectory. In one simulation step of the modified KMC algorithm, one first generates the next transition's time $t'=t^*$ according to \cref{eq:marg_t} and then generates the next state $x'$ according to \cref{eq:Pcond_x} given the proposed time $t'=t^*$. One can verify that the generated pair of $x'$ and $t'$ follows the true single-step trajectory probability distribution (\cref{eqn:trajectory}). This analysis is applicable to arbitrary time-dependent processes, where the system's rates change over time either due to (1) open-loop control according to a scheduled protocol\cite{Anderson2007-dq,Anderson2015-ql, Anderson2011-gc, Chow1996-ab, Li2009-ee} or (2) closed-loop control with a series of feedback measurements. 


\subsection{Review of open-loop-control KMC: General algorithm}\label{subsec:nonfeedback}
In this section, we review the open-loop-control KMC methods previously described in \cite{Anderson2007-dq,Anderson2015-ql, Anderson2011-gc, Chow1996-ab, Li2009-ee}. This review also serves as an alternative derivation and justification of these algorithms in the language of single-step trajectory generation introduced in \cref{subsec: single-step}. The representation of the single-step trajectory probability as the product of a marginal probability and a conditional probability, \cref{eq:joint_x't'},  allows one to separately generate the next transition's time $t'$ and state $x'$, even when the time-dependence of the system's rate creates statistical correlation between $t'$ and $x'$. 

{\it Generating transition time $t'$:}
The random generation of transition time $t'>t_k$ according to the marginal probability density in \cref{eq:marg_t} can be realized by first obtaining the cumulative density function
\begin{equation}
    \text{c.d.f.}(t') = 1- e^{\int_{t_k}^{t'} R_{xx}(t) {\rm d}t}
\end{equation}
and then generating a random number $u$ following a uniform distribution between $0$ and $1$, and finally solving for $t^*$ according to the following equation:
\begin{equation}
    u = \text{c.d.f.}(t^*) = 1- e^{\int_{t_k}^{t^*} R_{xx}(t) {\rm d}t}
\end{equation}
Or equivalently, for a random number $v=1-u$ also following a uniform distribution between $0$ and $1$, one can solve $t^*$ as the root of the following equation
\begin{equation}
\label{eqn:rxx}
    \ln (v) = H(t^*) ~,~ t^*\in (t_k, \infty)
\end{equation}
where
\begin{equation}
\label{eq:Ht*}
    H(t^*) = \int_{t_k}^{t^*} R_{xx}(t) {\rm d}t  ~,~ t^*\in (t_k, \infty)
\end{equation}
and $R_{xx}(t)$ is the negative escape rate from state $x$ at arbitrary time $t$. 

{\it Generating the next state $x'$:}
Once the transition time $t^*$ is generated, the selection of the next state $x'$ is straightforward -- the new state's probability is proportional to their transient transition rates evaluated at the chosen transition time $t^*$, as defined in \cref{eq:Pcond_x}. This resembles the state generation in the direct KMC method, with the only difference being the probability to jump to state $x'$ here is proportional to the transient rate $R_{x'x_k}(t^*)$ evaluated at the chosen time $t^*$. 

To summarize, one simulation step of the open-loop-control KMC algorithm can be enumerated in \cref{alg:non-feedback}. Furthermore, by including appropriate approximations (see \cref{subsec:piece}), one can reproduce various open-loop-control KMC algorithms previously described in \cite{Anderson2007-dq,Anderson2015-ql, Anderson2011-gc, Chow1996-ab, Li2009-ee}.

\begin{algorithm}
    \caption{Open-loop-control KMC}
    \label{alg:non-feedback}
    \begin{algorithmic}[1]
        \STATE The starting time of the current simulation step is $t$ with the starting state $x$.
        \STATE Generate the list of all possible states that the system can jump to, and evaluate the time-dependent escape rate from state $x$ as $-R_{xx}= \sum_{x'\neq x} R_{x'x}(t)$ where the summation takes all possible new states $x'$ into consideration.
        \STATE Generate a random number $v \in (0,1]$ following a uniform distribution. Obtain the next event time $t^*$ by solving for $t^*>t$ from \cref{eqn:rxx}.
        \STATE Given the proposed transition time $t^*$, propose the transition event (e.g., to state $x'$) with the probability proportional to their transient transition rates $p(x') \propto R_{x'x}(t^*)$ (\cref{eq:Pcond_x}). 
        \STATE Update the system state to $x'$ and time to $t^*$. Continue the iteration by going to step 2.
    \end{algorithmic}
\end{algorithm}

\subsection{Piece-wise approximations to open-loop control}
\label{subsec:piece}
In practice, solving for the next event's time $t^*$ by a given random number $v$ and the equation \cref{eqn:rxx} for arbitrarily complex time-dependent rates may be challenging. We now briefly review a few ways to approximately solve for this integral equation. Various methods have been proposed in simulating specific biological processes.\cite{Anderson2015-ql,Chow1996-ab, Ding2016-qr, Li2009-ee} For illustrative purposes, we demonstrate two approximations and sketch the corresponding algorithms for: (i) piece-wise linear function approximation and (ii) piece-wise step function approximation for the escape rate $-R_{xx}(t)$. 

\begin{figure}[h]
    \includegraphics{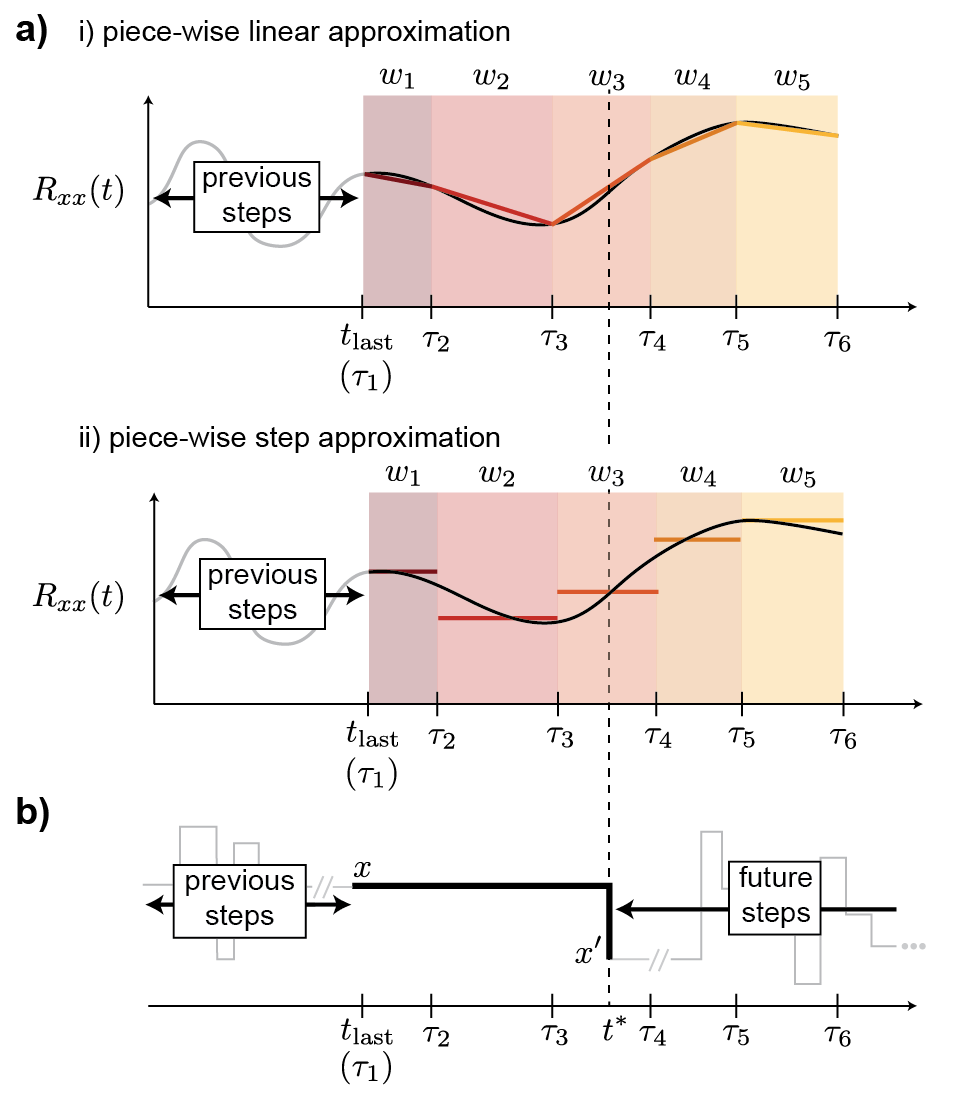}
    \caption{Approximating open-loop protocol control. a) Graphical representation of both (i) piece-wise linear rate function and ii) piece-wise step-function rates. The black curve is the actual negative escape rate $R_xx(t)$. The approximation breaks the time into multiple windows. Each window $w_i$ starting at time $\tau_i$ is represented by shaded regions, and the colored straight lines represent the approximated piece-wise negative escape rate. b) Cartoon of one possible trajectory of the system. Here the trajectory within the step of simulation is shown by the bold black lines; previous and future trajectory are shown by light gray lines. This simulation step is conditioned on the starting time $t_{\rm last}$ and state $x$, and proposes to jumped to a new state $x'$ at time $t^*>t_k$.}
    \label{fig:non-feedback}
\end{figure}

Here we illustrate the algorithms by focusing on one step of the simulation. Let us denote the previous state as $x_k=x$ and the previous time $t_k = t_{\rm last}$; then the algorithm's task is to generate the next event's time $t^*$ and the next state $x'$. As illustrated in \cref{fig:non-feedback}a, the negative escape rate from the state $x$, $R_{xx}(t)$ (shown as black curve) can be approximated by (i) a consecutive set of straight line segments (top panel), or (ii) approximated by a set of step-wise step functions (bottom panel).

(i) \textit{Piece-wise linear escape rates.}
Let us denote each window $w_i$ by a time interval $(\tau_i,\tau_{i+1})$ and also set the beginning of the first window at $\tau_1=t_{\rm last}$. The negative escape rate from state $x$ at each window's left edge is denoted by $R_{xx}^{\tau_i} = - \sum_{x'\neq x} R_{x'x}(\tau_i)$. Under this assumption, the numerical integration involved in \cref{eqn:rxx,eq:Ht*} can be considered as:
\begin{equation}
\label{eq:11}
    H(t^*) = A_1+ A_2+ \cdots A_{i-1} + \int_{\tau_{\tau_i}}^{t^*} R_{xx}(t) {\rm d} t
\end{equation}
for $\tau_i < t^* <\tau_{i+1}$. Here the integral for each window starting from $\tau_1 =t_{\rm last}$ is defined as
\begin{equation}
\label{eqn:piecewise linear area}
    A_i = (\tau_{i+1} - \tau_{i})R_{xx}^{\tau_{i}} + \frac{R_{xx}^{\tau_{i+1}} - R_{xx}^{\tau_{i}}}{2} (\tau_{i+1} - \tau_{i})
\end{equation}
Here, $A_i$ is the integrated negative escape rate within the window $w_i$ defined from $\tau_i$ to $\tau_{i+1}$. Successive windows $w_i$ starting from $i=1$ will be evaluated until the $j$-th window where the transition time is chosen, which can be identified by $\sum_{i=1}^{j-1} A_i < -\ln v < \sum_{i=1}^j A_i$. Given \cref{eq:11,eqn:piecewise linear area}, we can solve for the $t^*$ according to \cref{eqn:rxx} in sequence, starting from $i = 1$ and continuing until the appropriate $t^*$ can be drawn from the $j-$th window according to:\footnote{In practical implementation of the numerical code, when $q_i \approx q_{i-1}$, the quadratic equation reduces to the linear, which can be solved from \cref{eqn:piecewise t*}.} 
\begin{equation}
    \sum_{i=1}^{j-1} A_i + R_{xx}^{\tau_{j}} (t^*-\tau_{j}) + \frac{R_{xx}^{\tau_{j+1}} - R_{xx}^{\tau_{j}}}{2 (\tau_{j+1} - \tau_{j})} (t^*-\tau_{j})^2 = \ln v 
\end{equation}

(ii) \textit{Piece-wise step-function escape rates.}
When the time dependent escape rate from state $x$ is considered as a piece-wise step function, where $R_{xx}(t)=R_{xx}^{\tau_i}$ for each window $t\in (\tau_i,\tau_{i+1})$, we can determine the next event's time $t^*$ by solving for a linear equation
\begin{equation}
\label{eqn:piecewise t*}
    (t^* -\tau_k)R_{xx}^{\tau_k} + \sum_{i=1}^{k-1} B_i = \ln v
\end{equation}
where $v$ is a random number uniformly distributed between $0$ and $1$ and the integrated negative escape rate for each window is denoted by
\begin{equation}
\label{eqn:piecewise step area}
    B_i = (\tau_{i+1} - \tau_{i}) R_{xx}^{\tau_i}
\end{equation}
By solving for the above equations starting from the first window, one can find the next event's time as $t^*\in (\tau_k,\tau_{k+1})$ occurring at $k$-th window. We demonstrate this method in a 1D lattice diffusion problem as shown in the Appendix.

\subsection{Closed-loop feedback controlled systems: General algorithm}
\label{subsec:feedback control}
Consider a feedback control scheme where a scheduled series of measurements will be taken at a set of observation time points $\tau_1, \cdots, \tau_i,\tau_{i+1},\cdots \tau_{i+k},\cdots$ (see \cref{fig:feedback}a). At each observation time $\tau_j$, the measurement outcome is determined by the transient state of the stochastic system: $m_j=x(\tau_j)$.\footnote{Here the measurement of the system may not be as detailed as the specific state that the system is in. It is possible that a subset of states will give a same measurement outcome (i.e., coarse-grained measurement). The method can be applied to both cases.}  Immediately, the measurement outcome updates the system's kinetic rates via a rapid feedback control mechanism. Let us denote the control signal as an explicit function of time $\lambda(t)$, which can be represented by a piece-wise function where each piece is defined for the time between two consecutive measurements and is determined by all previous measurement outcomes. The actual protocol can be piece-wise constant (see illustrative example in \cref{subsec:application,subsec: refrigerator}), or in more general cases, could be a solution of an ODE parameterized by the measurement outcomes at each feedback measurement. The general algorithm proposed here is applicable to arbitrary functional forms of the control protocol.

The key to correctly simulate the feedback controlled system is to recognize that, when carrying out each step of a KMC simulation, one only needs to predict the next transition event given the present state and the transition rates set by the control protocol.
Now consider a system placed at state $x$ and time $t_{\rm last}$ set by the previous step of simulation. Without losing generality, let us consider the current time as $\tau_{i-1}<t_{\rm last}<\tau_i$, so that the current system has gone through the $(i-1)-$th measurement and the very next feedback measurement (the $i-$th measurement) will occur at a future time $\tau_i$. 

To carry out the present simulation step and predict the next transition time $t^*$ and next state $x'$, there are only two possible scenarios to consider, which are separated by a {\it milestone time} -- the immediate next measurement time, $\tau_i$. The two scenarios are the following: (i) the next event takes place at a time before the milestone time ($t^* < \tau_{i}$, \cref{fig:feedback}b[i]) or (ii) the next event takes place at any time after the milestone time ($t^* >\tau_i$, \cref{fig:feedback}b[ii]). The milestone time $\tau_i$ is highlighted by the color scheme transition from red to yellow in \cref{fig:feedback}b[ii]. This color scheme is also used in \cref{eqn:determined lambda} and \cref{eqn:projected lambda}.

In the first case, one does not need to consider the control protocol after the next measurement and only needs to consider the protocol before $\tau_i$. For this case, the control protocol for any $t<\tau_i$ is fully determined without any ambiguity, given the historical trajectory of the simulation.
This {\it pre-milestone} part of control protocol before $\tau_i$ can be generally represented by the following function and is fully determined:
\begin{equation}
\label{eqn:determined lambda}
    \lambda(t)  =
    \begin{cases}
        \lambda_1 \left(t; m_1\right) & \tau_1 \leq t < \tau_2 \\
        \vdots \\
        {\color{Mahogany} \lambda_{i-1}(t; m_1, \cdots, m_{i-1})} & {\color{Mahogany} \tau_{i-1} \leq t < \tau_{i}} 
    \end{cases}
\end{equation}
where $\lambda_j$ denotes the control protocol for time window $\tau_j < t \leq \tau_{j+1}$, and its functional form can be impacted by the all previous measurement results $m_1, m_2, \cdots, m_j$ through any prescribed feedback mechanisms. Here, $j$ must be smaller than $i$.


In the second case, we need to treat the system with a step-wise protocol where the system still evolves according to the pre-milestone protocol (\cref{eqn:determined lambda}) up to the milestone time $\tau_i$, and then the protocol is updated to a future post-milestone protocol. It may appear that the post-milestone protocols (for $t>\tau_i$) cannot be fully determined due to the ambiguity of the future system states. However, as we have argued, each step of the KMC simulation involves the generation of a single step trajectory that starts at $(x,t_{\rm last})$, remains at state $x$ and only jumps to an unknown state $x'$ at a future unknown transition time $t^*$. Thus, for any unknown transition time $t^*>\tau_i$ larger than the milestone time, each futuristic feedback measurement according to the trajectory under consideration will return a measurement outcome based on the system state $x$. 
In other words, the post-milestone protocol for any future $t \geq \tau_{i}$ is denoted by:
\begin{equation}
\label{eqn:projected lambda}
    {\color{mydarkorange} \tilde \lambda(t)} = 
    \begin{cases}
            {\color{mydarkorange} \tilde\lambda_{i} (t; m_1, \cdots, m_{i-1}, x)} & {\color{mydarkorange} \tau_i \leq t < \tau_{i+1}} \\
            {\color{mydarkorange} \tilde\lambda_{i+1} (t; m_1, \cdots, m_{i-1}, x, x)} & {\color{mydarkorange}  \tau_{i+1} \leq t < \tau_{i+2}} \\
            {\color{mydarkorange} \tilde\lambda_{i+1} (t; m_1, \cdots, m_{i-1}, x, x, x)} & {\color{mydarkorange}  \tau_{i+2} \leq t < \tau_{i+3}} \\
            {\color{mydarkorange} \vdots}
    \end{cases}
\end{equation}
which is determined without ambiguity because all future measurements must detect the system at the current state $x$ (i.e., $m_i, m_{i+1},m_{i+2},\cdots =x$). In this argument, all futuristic protocols are then determined by the combination of the historical measurement outcomes and the asserted futuristic measurement outcomes, $m_1, m_2, \cdots,m_{i-2}, {\color{Mahogany}m_{i-1}}, {\color{mydarkorange} x,x,x,\cdots}$.


\begin{algorithm}
    \caption{Closed-loop-control KMC}\label{alg:feedback}
    \begin{algorithmic}[1]
        \STATE In the current simulation step, the last transition event takes place between the $(i-1)$-th and $i$-th measurement $\tau_{i-1}<t<\tau_i$, and landing at the present state $x$; the control protocol $\lambda(t)$ can be determined up to $t<\tau_i$, (see \cref{eqn:determined lambda}) with the last measurement outcome being $m_{i-1}=x(\tau_{i-1})$. (See \cref{fig:feedback})
        \STATE To project future control protocols for $t\geq\tau_i$, take the current state $x$ for all future measurements. The projected control protocols are written in \cref{eqn:projected lambda}.   
        \STATE Generate the list of all possible states that the system can jump to from current state $x$, and evaluate the time-dependent escape rate from state $x$ as $-R_{xx}(t)= \sum_{x'\neq x} R_{x'x}(t)$. This time-dependent rates $R_{x'x}((\lambda(t)))$ are determined by the protocol $\lambda(t)$. The control protocol are described by the combination of the deterministic part for $t<\tau_i$ \cref{eqn:determined lambda} and the projected part for $t\geq \tau_i$ \cref{eqn:determined lambda,eqn:projected lambda}.
        \STATE Generate a random number $v \in (0,1]$ following a uniform distribution. Obtain the next event time $t^*$ by solving for $t^*>t$ from \cref{eqn:rxx}, where $H(t^*)$ is defined by \cref{eq:Ht*} where the rates $R_{xx}(\lambda(t))$ are dictated by the protocols defined in \cref{eqn:determined lambda,eqn:projected lambda}. 
        \STATE  Given the proposed transition time $t^*$, propose the transition event (e.g., to state $x'$) with the probability proportional to their transient transition rates (\cref{eq:Pcond_x}). If $t^* <\tau_i$, the transient rates are determined by the deterministic protocol from \cref{eqn:determined lambda}. If $t^* >\tau_i$, the rates are determined by the projected protocol from \cref{eqn:projected lambda}.
        \STATE Update the system time to $t^*$ and state to $x'$. If the time duration between $t$ and $t^*$ includes one or more scheduled feedback measurements, also record measurement results of state $x$. 
        \STATE Continue the iteration by returning to step 2.
    \end{algorithmic}
\end{algorithm}

In summary, through our single-step trajectory probability argument introduced in \cref{subsec: single-step}, we are able to assert a fully determined protocol for a feedback control system. This determined protocol combines \cref{eqn:determined lambda}, the pre-milestone protocol $\lambda(t)$ for $t<\tau_i$, which is fully determined by historical feedback measurements, and \cref{eqn:projected lambda}, the futuristic post-milestone protocol $\tilde \lambda(t)$ for $t>\tau_i$, which appears to be unknown due to the future feedback measurement outcomes. As a result, even for closed-loop controlled systems, one can utilize an open-loop-control KMC algorithm\cite{Anderson2007-dq,Anderson2015-ql, Anderson2011-gc, Chow1996-ab, Li2009-ee} reviewed in \cref{subsec:nonfeedback} to generate the next transition event according to the correct probability distribution \cref{eq:joint_x't',eq:marg_t,eq:Pcond_x}. This general algorithm is shown in \cref{alg:feedback}. For illustrative purpose, the following demonstrates a simple application of the proposed feedback-control KMC algorithm.

\subsection{Application: Piece-wise-constant feedback control}
\label{subsec:application}
Here, we demonstrate the proposed new algorithm with a general yet simple scenario. In this example, the rates within each measurement window remain constant and are instantaneously updated based on each measurement. For more complicated feedback control regimes (e.g., the control protocol evolves deterministically in time according to differential equations parameterized by historical measurement outcomes), one can simply extend this algorithm in accordance with general feedback controlled KMC algorithm described in \cref{alg:feedback}.

\begin{figure}[h]
    \centering
    \includegraphics{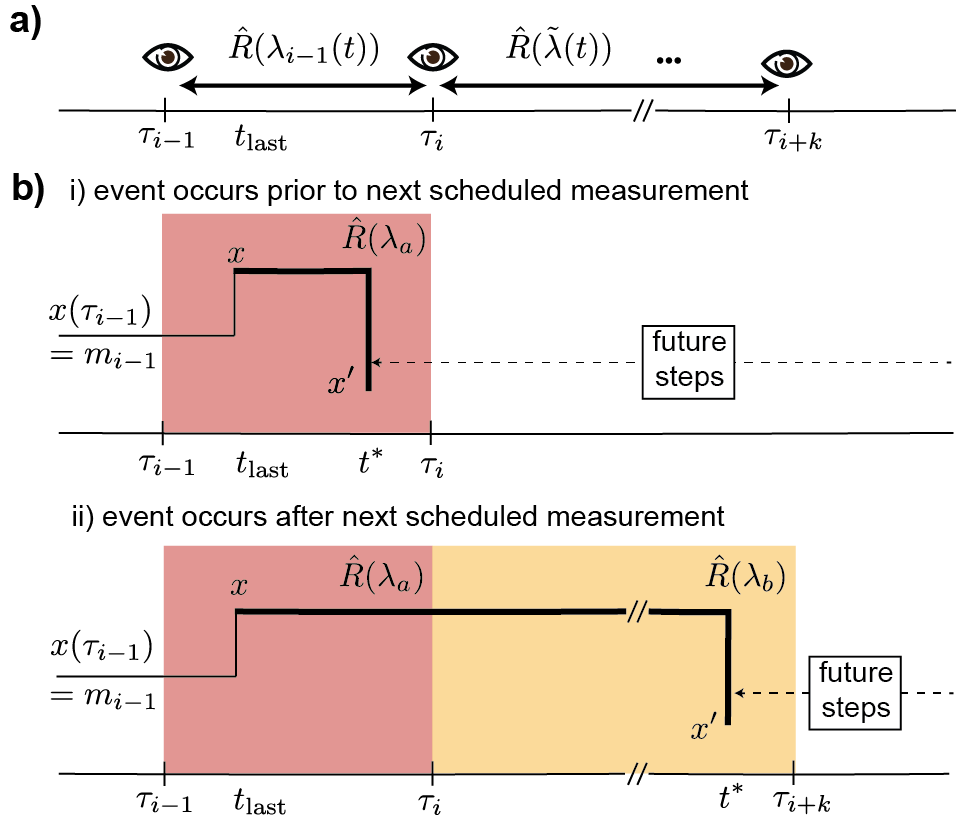}
    \caption{Closed-loop control protocol. a) Graphical representation of a series of scheduled measurements occurring at $\tau_{i-1}, \tau, \cdots, \tau_{i+k}$. b) Two types of single-step trajectories illustrated in the piece-wise step-function feedback control: i) the next event occurs prior the the next scheduled measurement evolved using the determined protocol  based on the last measurement $R_{xx(\lambda_a})$ or ii) event occurs after the $k$ additional measurements, and the rate is projected by fixing the future measurements to be state $x$ resulting in control signal being $\lambda_b$, and the escape rate is $R_{xx}(\lambda_b)$.}
    \label{fig:feedback}
\end{figure}

We here describe a model algorithm where the control parameter $\lambda(t)$ between two consecutive measurement events are always maintained at a constant value. In this case, both the determined protocol $\lambda_{i-1}(t) =\lambda_a,~t_{\rm last}<t<\tau_i$ and the projected protocols $\tilde \lambda(t)=\lambda_b,~t>\tau_i$ are both piece-wise constants. Therefore, the escape rate from state $x$ is a piece-wise constant function with only two pieces (see \cref{fig:feedback}b):
\begin{equation}
\label{eqn:piecewise rate}
    R_{xx }(\lambda(t)) =
    \begin{cases}
        R_{x x} (\lambda_a)  &~ t_{\rm{last}} \leq t < \tau_i \\
        R_{x x} (\lambda_b) & ~t \geq \tau_i
    \end{cases}
\end{equation}

In this case, the simulation chooses the next event's occurrence time by solving for the solution of the piece-wise function:
\begin{equation}
\label{eqn:piecewise rxx}
    \ln (v) =
    \begin{cases}
        R_{x x} (\lambda_a) (t^* - t_{\rm last}) & t < \tau_i \\
        R_{xx}(\lambda_b) (t^* - \tau_i) + R_{xx}(\lambda_a)  (\tau_i - t_{\rm last}) & t \geq \tau_i
    \end{cases}
\end{equation}
where $v$ is a uniformly distributed random number between $0$ and $1$, and the $r.h.s$ of the equation is defined by \cref{eq:Ht*}. 
There are two types of possible simulation outcomes of the single-step trajectories. At possibility $\#1$ (top panel of \cref{fig:feedback}b), when the next event occurs before the next measurement, $t_{\rm{last}}< t^* < \tau_{i}$, the dynamical rates remains unchanged throughout the time, and the rates was determined by the last measurement outcome $m_{i-1}=x(\tau_{i-1})$ (see red shaded regions in \cref{fig:feedback}b). At possibility $\#2$ (bottom panel of \cref{fig:feedback}b), when the next event occurs after $k$ more future measurements $\tau_{i+k-1}< t^* < \tau_{i+k}$, for $k=1,2,\cdots$, the dynamics are initially governed by the old rates $\hat R(\lambda_a)$ before the $i$-th measurement, and then by a new rate $\hat R(\lambda_b)$ after $\tau_i$ (see yellow shaded regions in \cref{fig:feedback}b). We demonstrate this method in a two-state refrigerator model in \cref{subsec: refrigerator}.

\section{Results and Discussion}\label{sec: results}
\subsection{Demonstration: 2-state feedback control Maxwell's Refrigerator}\label{subsec: refrigerator}
To demonstrate the new feedback controlled KMC algorithm proposed in \cref{subsec:feedback control}, we utilize an example feedback control system whose behavior can be solved analytically. Feedback control systems have been extensively studies in various stochastic systems\cite{bechhoefer2021control, annby2022quantum, horowitz2014quantum, horowitz2011thermodynamic, horowitz2011designing, horowitz2010nonequilibrium, bhattacharyya2022feedback, sagawa2012nonequilibrium,Sagawa2010-zv,sagawa2012fluctuation,parrondo2015thermodynamics}. Here we choose to simulate the stochastic trajectories of a 2-state refrigerator inspired by reference\cite{mandal2013maxwell}. 

\begin{figure}[h]
    \centering
    \includegraphics{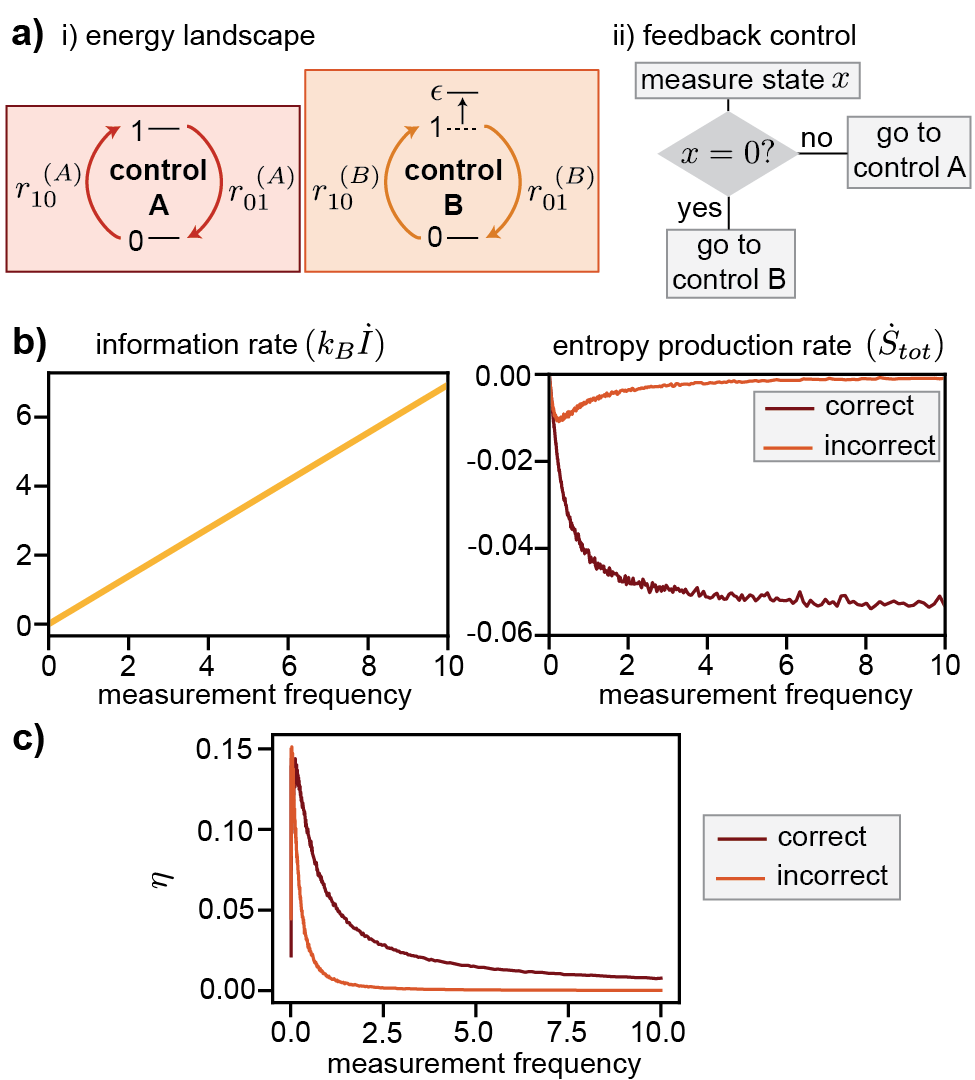}
    \caption{Two-state refrigerator through periodic feedback control. a) i) Design of energy landscape of two-state system in two different conditions and ii) feedback control decision occurring at each measurement time. b) Time-averaged thermodynamic properties of the refrigerator including (left) information rate and (right) entropy production rate computed under different feedback measurement frequencies. To demonstrate the artificial Zeno effect, both the correct feedback controlled KMC method (correct) and an naively implemented direct KMC method (incorrect) are shown. c) System's thermodynamic efficiency ($\eta\leq 1$) calculated under different feedback control frequencies with both the correct and the incorrect algorithms.}
    \label{fig:refrigerator}
\end{figure}

Consider a 2-level system coupled to a heat bath. The ground state level's energy is $E_0=0$ and the excited level's energy is controlled externally by a binary toggle switch. At control $A$ the excited state energy is $E_1^A=1$; at control condition $B$, $E_1^B=\epsilon >1$. The feedback control is facilitated by a periodic sequence of instantaneous measurements to the system's state at times $\tau_i=i/\nu$, where $\nu$ is the measurement frequency. If the measurement detects the system at the ground state ``0'', the control is switched to $B$, and if excited state ``1'', control is $A$(see \cref{fig:refrigerator}a).

This system falls into the category of the step-wise constant feedback control scenario (discussed in \ref{subsec:application}). In this case, the system evolves either according to the transition rates set by control condition $A$ or by condition $B$. At each instantaneous time, the transition rates takes the Arrhenius form $R_{ij} = e^{B_{ij} - E_j}$, where we have taken the unit such that the inverse temperature $\beta=1/(k_BT)=1$. In our demonstration, we set the barrier for the transition to $B_{ij}=2$, and set the energies according to $E_1^A=1$, and $E_1^B=\epsilon =1.5$. Under this feedback control, it was shown that the system behaves as a refrigerator, constantly drawing energy from the heat bath and performs work against the external control. This refrigerator does not require energy expenditure but rather operates at the cost of information obtained via the measurements. For a comprehensive overview of the definition of heat and work in the stochastic systems, refer to the book \cite{Sekimoto_undated-xn}. For the connection between information and thermodynamic entropy, a few representative works can be found in \cite{penrose2005foundations, bennett1973logical,landauer1961irreversibility, mandal2012work,zurek1989thermodynamic,deffner2013information, strasberg2013thermodynamics, sagawa2012nonequilibrium,Sagawa2010-zv,sagawa2012fluctuation,parrondo2015thermodynamics, lu2014engineering}.

At each stochastic transition from the ground state to the excited state, the system draws from the heat bath energy $Q_{\rm sys}=1$ (for control $A$) or $Q_{\rm sys}=\epsilon$ (for control $B$). At the transition from excited state to the ground state, the system's heat gain is $Q_{\rm sys}=-1$ (for control $A$) and $Q_{\rm sys}=-\epsilon$ (for control $B$). At any schedule measurement causing a control switch from $A$ to $B$ (the system is at the ground state), there is no work or heat exchange ($W=0$, $Q=0$). At the scheduled measurement causing a control switch from $B$ to $A$, the system performs positive work to the controller $W_{\rm ext}=\epsilon-1$, and there is no heat exchange $Q=0$ within the instantaneous measurement time. By counting the frequencies of the stochastic transitions of system ($0$ to $1$ or $1$ to $0$) under different control conditions $A$ and $B$ respectively, one can calculate the average heat rate (the average heat drawn by the system per unit time). 

By implementing the feedback controlled KMC algorithm, we are able to obtain an ensemble of 100 statistically independent stochastic trajectories from such a feedback controlled refrigerator. Each trajectory is of time length 500 and we exclude the initial relaxation period of length 100 from statistics. From the trajectory, one can compute the bath's entropy change rate, which is equal to the total entropy change rate of the system and the bath:
\begin{equation}
     \dot S_{tot} = \frac{J^A_{0\rightarrow 1}\cdot (E_0-E_1^A)  +j^B_{0\rightarrow 1}\cdot (E_0-E_1^B)}{T}
\end{equation}
where the net transition current $J$ is defined by the difference between the detailed currents:
\begin{align}
    J^A_{0\rightarrow 1}&=j^A_{0\rightarrow 1}-j^A_{1\rightarrow 0}\\
    J^B_{0\rightarrow 1}&=j^B_{0\rightarrow 1}-j^B_{1\rightarrow 0}
\end{align}
Furthermore, since each measurement is binary, one can calculate the information rate simply by multiplying the information per measurement with the measurement frequency: $k_B \dot I = \nu \ln 2$. 
Finally, we also computed the efficiency of the refrigerator as the entropy decrease rate over the information gain rate:
\begin{equation}
    \eta = \frac{-\dot S_{tot} }{k_B \dot I} \leq 1
\end{equation}
which characterizes the system's ability to decrease entropy at the expenditure of information change. Notice that according to the generalized second law of thermodynamics $\dot S_{tot} + k_B \dot I \geq 0$,\cite{sagawa2012nonequilibrium,sagawa2012fluctuation,Sagawa2010-zv,parrondo2015thermodynamics}, the efficiency must be lower than or equal to 1. The simulation results for a range of measurement frequencies are shown in \cref{fig:refrigerator}b-c.

One can verify the proposed feedback controlled KMC by comparing its result to the analytical solution of the system. In the limit of the infinitely high frequency measurements, $\nu\gg 1$, one can obtain an effective constant rate matrix of the system and analytically obtain its average heat rate and entropy change rate. 
Due to infinitely frequent measurements, the system's transition rate from state $1$ to $0$ is always equal to $R^{\rm eff}_{01}=R^A_{01}=e^{-1}$ and the system's transition rate from state $0$ to $1$ is always equal to $R^{\rm eff}_{10}=R^B_{10}=e^{-2}$. As a result, the effective rate matrix under infinite frequency feedback control is
\begin{equation}
    \hat R^{\rm eff} = 
    \begin{bmatrix}
        -e^{-2} & e^{-1} \\
        e^{-2} & -e^{-1}
    \end{bmatrix}
\end{equation}
The steady state probability for this system can be written as $(p^{\rm ss}_0,p^{\rm ss}_1) =(\frac{e}{1 + e}, \frac{1}{1 + e})$. This allows us to compute the detailed transition current from 0 to 1 as $\frac{e}{1 + e} e^{-2}$, and current from 1 to 0 as $\frac{1}{1 + e} e^{-1}$. Due to the frequent measurement, each transition from 0 to 1 must occur under control $B$, which takes energy $\epsilon$ from the bath, and each transition from 1 to 0 must occurs under control $A$, which deposits energy $1$ into the bath. As a result, for a system at the steady state (no system entropy change), the total entropy change of the system and the bath is purely the bath entropy change with the rate
\begin{equation}
\label{eq:result}
     \lim_{\nu\rightarrow \infty}\dot S_{tot} = \frac{j^A_{1\rightarrow 0}\cdot 1  + j^B_{0\rightarrow 1}\cdot (-\epsilon)}{T} \approx -0.0494690099
\end{equation}
where the detailed steady-state currents are
\begin{align}
    j^A_{1\rightarrow 0}&= R^{\rm eff}_{01}\cdot  p^{\rm ss}_1 =R^{A}_{01}\cdot  p^{\rm ss}_1 \\
    j^B_{0\rightarrow 1}&=R^{\rm eff}_{10} \cdot p^{\rm ss}_0=R^{B}_{10} \cdot p^{\rm ss}_0
\end{align}
From \cref{fig:refrigerator} one can verify that the entropy change rate at high frequency limit obtained from our proposed feedback controlled KMC algorithm matches well with the analytical result in \cref{eq:result}

\subsection{Discussion: Avoiding artificial Zeno effect}\label{subsec: lattice diffusion}
This work proposed a new algorithm to simulate systems under feedback control. Here we illustrate that for systems with time-dependent control (both open-loop and closed-loop), the direct implementation of the original Gillespie algorithm\cite{gillespie2001approximate} can lead to erroneous results, where the system's dynamics appears to be immobilized due to frequent changes of the external control parameter or frequent feedback controls. We name this erroneous results the  {\it artificial Zeno effect}. This effect appears to resemble the quantum Zeno effect\cite{Fischer2001-bn} but is fundamentally different -- the former is an artificial numerical error, and the latter is a true physical phenomenon.

In the quantum Zeno effect\cite{Fischer2001-bn} frequent projective measurements may reset the system's wave function into an eigenstate and, in the limit of infinitely high measurement frequency, the system's wave function is frozen at the eigenstate and the system appears to be non-evolving. In contrast, classical stochastic systems, where the measurement does not directly impact the system's state\footnote{Under feedback control measurements, the system's kinetic rates can be updated after the measurement, but the system's state is not directly altered by the action of measurement.}, should not have any Zeno effect. The artificial Zeno effect is a numerical error only when one naively implements the direct KMC algorithm in a piece-wise manner for time-dependent systems.

For illustrative purpose we first demonstrate the erroneous artificial Zeno effect in the feedback control system discussed in \cref{fig:refrigerator}. In the Appendix, this discussion has been extended to an open-loop controlled time-dependent systems (\cref{fig:diffusion}). 

For the feedback controlled system, one naive way to implement the original Gillespie algorithm\cite{gillespie2001approximate} is to reset the KMC simulation at each measurement time point. Specifically, one may perform a constant-rate KMC simulation for each measurement time window. Consider a system at state $x$ at time $<\tau_{i-1}<t<\tau_i$. If the direct KMC approach proposes the next transition's time to be $t^* >\tau_i$, one simply evolves the system time to $t' =\tau_i$ while maintaining the system state to be $x$ and then starts another step of the direct KMC simulation, with the initial state $x$ and initial time $\tau_i$. This approach appears to be valid, since the predicted time $t^*$ is indeed longer than the time of the next measurement, and the system should not change until after the next measurement. However, this direct implementation of the KMC method leads to wrong results that can become obvious by considering the infinitely frequent measurement limit, where the time till the next measurement approaches $0$ and the chance that the direct KMC proposes a $t^*$ before the next measurement becomes zero. Thus the simulation is frozen to state $x$. This artificial Zeno effect becomes more prominent for higher frequency measurements. 

To illustrate this mistake, we implement the above direct KMC simulation for the Maxwell's refrigerator discussed in \cref{subsec: refrigerator} and show that the artificial Zeno effect significantly underestimates the transition events of the systems dynamics, resulting in an underestimation of the refrigerator's entropy decreasing rate and thermodynamic efficiency (see \cref{fig:refrigerator}bc).

For a feedback controlled system where one measures and acquires system's states at a scheduled sequence of times, plainly restarting the clock of the KMC simulation from the measured state to simulate the dynamics of the system will lead to an artificial Zeno effect. The closed-loop-control algorithm demonstrated here avoids this pitfall and preserves the correct dynamics. In the Appendix, we show that similar mistakes may occur in open-loop-control systems only if one overlooks previously introduced algorithms\cite{Anderson2007-dq,Anderson2015-ql, Anderson2011-gc, Chow1996-ab, Li2009-ee}

\section{Conclusion}
In conclusion, we here propose a new method to simulate stochastic trajectories of systems experiencing feedback control. This new method is derived based on a general single-step trajectory probability analysis. This analysis can be used to derive and recapitulate several previously proposed open-loop-control KMC algorithms for non-feedback control systems whose rates change in time under a given protocol. Then, we demonstrate that feedback controlled systems can be simulated by combining pre-milestone control protocol and a post-milestone control protocol. This allows us to perform a time-dependent KMC simulation for feedback controlled systems with a modified open-loop-control KMC algorithm. 
We validate our claims using simulation of a two-state feedback control Maxwell's refrigerator. Our correct algorithm provides the true dynamics of the system. The simulation results are compared with the analytical results obtained in the infinitely frequent controlled scenarios, and the numerical results are in perfect agreement with the analytical result. Furthermore, the new modified KMC algorithms avoid the artificial Zeno effect, where a wrongly implementation of the direct KMC may lead to an erroneous underestimation of the frequency of transitions. 

The algorithms described herein can find broad applicability in a number of fields, particularly in biological systems which experience rapid fluctuations in local environments and frequent feedback control via interaction with surroundings. Multiple processes -- such as neurophysiology,\cite{seidler2004feedforward} transcriptional regulation,\cite{henninger2021rna} circadian clocks,\cite{brown2020dual} and metabolic regulation\cite{chaves2019dynamics} -- are examples of such systems beyond the scope of analytical solutions. A KMC algorithm to treat these feedback controlled systems would allow the relevant dynamics to be sampled and captured. 
Even though the example provided in \cref{fig:refrigerator} is a binary state model which is analytically solvable, the algorithm can be directly applied to complex systems that are not analytically solvable (see \cref{subsec:application}). If the system's control protocol follows complex rules, one can still apply the general algorithm by making necessary approximations on the control protocol similar to those sketched in \cref{subsec:piece}. While rapid feedback control or complex dynamics may pose challenges by increasing the computational cost of these simulations, our derivation provides a flexible path to tackling these problems. We believe that the relevant non-equilibrium behavior and thermodynamic quantities in a wide variety of feedback controlled chemical and biological system, such as reaction turnover rates, binding affinity, or assembly properties, can be accurately captured by using the new algorithm. 

\section{Appendix}
If one overlooks the existing open-loop-control KMC methods\cite{Anderson2007-dq,Anderson2015-ql, Anderson2011-gc, Chow1996-ab, Li2009-ee} and approximates the control protocol by piecewise functions, it is possible that an artificial Zeno effect similar to \cref{fig:refrigerator} appears. Below we demonstrate a comparison between the correct method\cite{Anderson2007-dq,Anderson2015-ql, Anderson2011-gc, Chow1996-ab, Li2009-ee} and the erroneous result. 

\begin{figure}[h]
    \centering
    \includegraphics{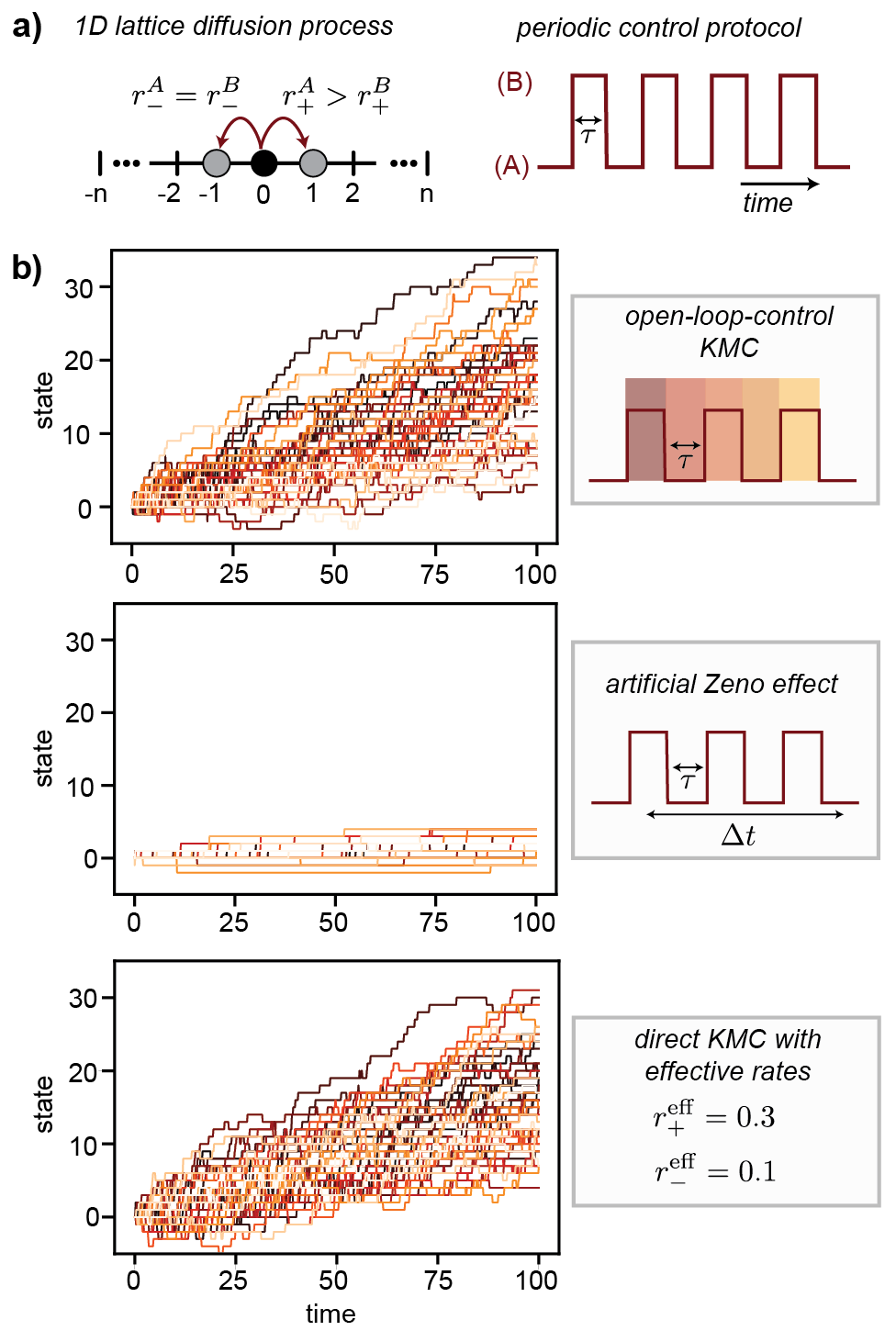}
    \caption{Diffusion along a 1D lattice. a) (left panel) Schematic of the diffusion processes with jumping rates to the right and left direction denoted as $r^{A/B}_+$ and $r^{A/B}_-$. The control condition is denoted by $A$ and $B$. (right panel) Periodic control protocol with period $2 \tau$. b) Simulated diffusion trajectories using (top panel) averaged rate dynamics in traditional KMC, (center panel) individual rate dynamics using incorrectly applied KMC due timescale mismatch between $\tau$ and time for each move $\Delta t$, (bottom panel) individual rate dynamics correctly simulated using the KMC described herein.}
    \label{fig:diffusion}
\end{figure}

Let us consider a diffusion process on a 1-dimension lattice, where random walkers experience biased transitions. 
The initial state of all random walkers are placed at position 0. Under a periodic switch between controls A and B, the system evolves under two sets of rates with control period set to $2\tau= 0.02$ (see \cref{fig:diffusion}a). In condition A, the transition rate for moving right is $r^A_+ = 0.4$ and for moving left is $r^A_- = 0.1$ (\cref{fig:diffusion}a, left panel). Conversely, in condition B, these rates are $r^B_+ = 0.2$ and $r^B_- = 0.1$. As the environment oscillates rapidly between conditions A and B, the system experiences alternating transition rates (\cref{fig:diffusion}a, right panel). To accurately simulate this scenario, we employ the correct open-loop-control KMC method to generate a set of trajectories representative of the random walker’s behavior (see \cref{fig:diffusion}b, top panel). To illustrate the erroneous artificial Zeno effect, we perform a piece-wise direct KMC simulation and find that the artificial Zeno effects significantly slowed down the system's diffusion, and the random walkers are mostly frozen without many transitions (see \cref{fig:diffusion}b, middle panel). In the end, to verify the validity of the open-loop-control KMC, we include an alternative way to obtain the true dynamics of random workers: by acknowledging that the control switch frequency is much greater than the diffusion rates, $\nu=1/(2\tau)=50 \gg r^{A/B}_{+/-}$, one can show that this system evolves under effective dynamics with constant rates \cite{tagliazucchi2014dissipative,zhang2023energy}. In this case, the system's dynamics can be generated by a direct KMC simulation under the effective rates, shown in \cref{fig:diffusion}b bottom panel. We verify that under the rapid oscillation limit, the effective dynamics and the dynamics generated by the open-loop-feedback KMC agrees with each other. However, the naive implementation of direct KMC to time-dependent dynamics results in a significant slow down of the dynamics (i.e., the artificial Zeno effect). 

\section*{Supplementary Information}
Numerical simulation codes can be found in the Supplementary Materials.

\section*{Data Availability Statement}
Data sharing is not applicable to this article as no new data were created or analyzed in this study.

\begin{acknowledgments}
This work was in part financially supported by the National Science Foundation under Award Number DMR-2145256. S.S.C acknowledges the National Science Foundation Graduate Research Fellowship (DGE-2040435).  We also thank the UNC Research Computing group for providing the computational resources and support towards these results. The authors are also grateful for valuable feedback from Prof. Chris Jarzynski, Dr. Hong Qian, Dr. Aaron Dinner, Dr. Zhongmin Zhang, and Mr. Jiming Zheng.
\end{acknowledgments}

\section*{References}
\bibliography{aipsamp}

\begin{thebibliography}{65}%
\makeatletter
\providecommand \@ifxundefined [1]{%
 \@ifx{#1\undefined}
}%
\providecommand \@ifnum [1]{%
 \ifnum #1\expandafter \@firstoftwo
 \else \expandafter \@secondoftwo
 \fi
}%
\providecommand \@ifx [1]{%
 \ifx #1\expandafter \@firstoftwo
 \else \expandafter \@secondoftwo
 \fi
}%
\providecommand \natexlab [1]{#1}%
\providecommand \enquote  [1]{``#1''}%
\providecommand \bibnamefont  [1]{#1}%
\providecommand \bibfnamefont [1]{#1}%
\providecommand \citenamefont [1]{#1}%
\providecommand \href@noop [0]{\@secondoftwo}%
\providecommand \href [0]{\begingroup \@sanitize@url \@href}%
\providecommand \@href[1]{\@@startlink{#1}\@@href}%
\providecommand \@@href[1]{\endgroup#1\@@endlink}%
\providecommand \@sanitize@url [0]{\catcode `\\12\catcode `\$12\catcode `\&12\catcode `\#12\catcode `\^12\catcode `\_12\catcode `\%12\relax}%
\providecommand \@@startlink[1]{}%
\providecommand \@@endlink[0]{}%
\providecommand \url  [0]{\begingroup\@sanitize@url \@url }%
\providecommand \@url [1]{\endgroup\@href {#1}{\urlprefix }}%
\providecommand \urlprefix  [0]{URL }%
\providecommand \Eprint [0]{\href }%
\providecommand \doibase [0]{http://dx.doi.org/}%
\providecommand \selectlanguage [0]{\@gobble}%
\providecommand \bibinfo  [0]{\@secondoftwo}%
\providecommand \bibfield  [0]{\@secondoftwo}%
\providecommand \translation [1]{[#1]}%
\providecommand \BibitemOpen [0]{}%
\providecommand \bibitemStop [0]{}%
\providecommand \bibitemNoStop [0]{.\EOS\space}%
\providecommand \EOS [0]{\spacefactor3000\relax}%
\providecommand \BibitemShut  [1]{\csname bibitem#1\endcsname}%
\let\auto@bib@innerbib\@empty
\bibitem [{\citenamefont {Shahrezaei}\ and\ \citenamefont {Swain}(2008)}]{shahrezaei2008stochastic}%
  \BibitemOpen
  \bibfield  {author} {\bibinfo {author} {\bibfnamefont {V.}~\bibnamefont {Shahrezaei}}\ and\ \bibinfo {author} {\bibfnamefont {P.~S.}\ \bibnamefont {Swain}},\ }\bibfield  {title} {\enquote {\bibinfo {title} {The stochastic nature of biochemical networks},}\ }\href@noop {} {\bibfield  {journal} {\bibinfo  {journal} {Current opinion in biotechnology}\ }\textbf {\bibinfo {volume} {19}},\ \bibinfo {pages} {369--374} (\bibinfo {year} {2008})}\BibitemShut {NoStop}%
\bibitem [{\citenamefont {Levine}\ and\ \citenamefont {Hwa}(2007)}]{levine2007stochastic}%
  \BibitemOpen
  \bibfield  {author} {\bibinfo {author} {\bibfnamefont {E.}~\bibnamefont {Levine}}\ and\ \bibinfo {author} {\bibfnamefont {T.}~\bibnamefont {Hwa}},\ }\bibfield  {title} {\enquote {\bibinfo {title} {Stochastic fluctuations in metabolic pathways},}\ }\href@noop {} {\bibfield  {journal} {\bibinfo  {journal} {Proceedings of the National Academy of Sciences}\ }\textbf {\bibinfo {volume} {104}},\ \bibinfo {pages} {9224--9229} (\bibinfo {year} {2007})}\BibitemShut {NoStop}%
\bibitem [{\citenamefont {Sz{\'e}kely~Jr}\ and\ \citenamefont {Burrage}(2014)}]{szekely2014stochastic}%
  \BibitemOpen
  \bibfield  {author} {\bibinfo {author} {\bibfnamefont {T.}~\bibnamefont {Sz{\'e}kely~Jr}}\ and\ \bibinfo {author} {\bibfnamefont {K.}~\bibnamefont {Burrage}},\ }\bibfield  {title} {\enquote {\bibinfo {title} {Stochastic simulation in systems biology},}\ }\href@noop {} {\bibfield  {journal} {\bibinfo  {journal} {Computational and structural biotechnology journal}\ }\textbf {\bibinfo {volume} {12}},\ \bibinfo {pages} {14--25} (\bibinfo {year} {2014})}\BibitemShut {NoStop}%
\bibitem [{\citenamefont {Earnshaw}\ and\ \citenamefont {Keener}(2010)}]{earnshaw2010invariant}%
  \BibitemOpen
  \bibfield  {author} {\bibinfo {author} {\bibfnamefont {B.~A.}\ \bibnamefont {Earnshaw}}\ and\ \bibinfo {author} {\bibfnamefont {J.~P.}\ \bibnamefont {Keener}},\ }\bibfield  {title} {\enquote {\bibinfo {title} {Invariant manifolds of binomial-like nonautonomous master equations},}\ }\href@noop {} {\bibfield  {journal} {\bibinfo  {journal} {SIAM Journal on Applied Dynamical Systems}\ }\textbf {\bibinfo {volume} {9}},\ \bibinfo {pages} {568--588} (\bibinfo {year} {2010})}\BibitemShut {NoStop}%
\bibitem [{\citenamefont {Lee}\ and\ \citenamefont {Othmer}(2010)}]{lee2010multi}%
  \BibitemOpen
  \bibfield  {author} {\bibinfo {author} {\bibfnamefont {C.~H.}\ \bibnamefont {Lee}}\ and\ \bibinfo {author} {\bibfnamefont {H.~G.}\ \bibnamefont {Othmer}},\ }\bibfield  {title} {\enquote {\bibinfo {title} {A multi-time-scale analysis of chemical reaction networks: I. deterministic systems},}\ }\href@noop {} {\bibfield  {journal} {\bibinfo  {journal} {Journal of mathematical biology}\ }\textbf {\bibinfo {volume} {60}},\ \bibinfo {pages} {387--450} (\bibinfo {year} {2010})}\BibitemShut {NoStop}%
\bibitem [{\citenamefont {Meister}\ \emph {et~al.}(2014)\citenamefont {Meister}, \citenamefont {Du}, \citenamefont {Li},\ and\ \citenamefont {Wong}}]{meister2014modeling}%
  \BibitemOpen
  \bibfield  {author} {\bibinfo {author} {\bibfnamefont {A.}~\bibnamefont {Meister}}, \bibinfo {author} {\bibfnamefont {C.}~\bibnamefont {Du}}, \bibinfo {author} {\bibfnamefont {Y.~H.}\ \bibnamefont {Li}}, \ and\ \bibinfo {author} {\bibfnamefont {W.~H.}\ \bibnamefont {Wong}},\ }\bibfield  {title} {\enquote {\bibinfo {title} {Modeling stochastic noise in gene regulatory systems},}\ }\href@noop {} {\bibfield  {journal} {\bibinfo  {journal} {Quantitative biology}\ }\textbf {\bibinfo {volume} {2}},\ \bibinfo {pages} {1--29} (\bibinfo {year} {2014})}\BibitemShut {NoStop}%
\bibitem [{\citenamefont {Anderson}\ and\ \citenamefont {Kurtz}(2011)}]{Anderson2011-gc}%
  \BibitemOpen
  \bibfield  {author} {\bibinfo {author} {\bibfnamefont {D.~F.}\ \bibnamefont {Anderson}}\ and\ \bibinfo {author} {\bibfnamefont {T.~G.}\ \bibnamefont {Kurtz}},\ }\bibfield  {title} {\enquote {\bibinfo {title} {Continuous time markov chain models for chemical reaction networks},}\ }in\ \href@noop {} {\emph {\bibinfo {booktitle} {Design and Analysis of Biomolecular Circuits: Engineering Approaches to Systems and Synthetic Biology}}},\ \bibinfo {editor} {edited by\ \bibinfo {editor} {\bibfnamefont {H.}~\bibnamefont {Koeppl}}, \bibinfo {editor} {\bibfnamefont {G.}~\bibnamefont {Setti}}, \bibinfo {editor} {\bibfnamefont {M.}~\bibnamefont {di~Bernardo}}, \ and\ \bibinfo {editor} {\bibfnamefont {D.}~\bibnamefont {Densmore}}}\ (\bibinfo  {publisher} {Springer New York},\ \bibinfo {address} {New York, NY},\ \bibinfo {year} {2011})\ pp.\ \bibinfo {pages} {3--42}\BibitemShut {NoStop}%
\bibitem [{\citenamefont {Bowman}, \citenamefont {Pande},\ and\ \citenamefont {No{\'e}}(2013)}]{Bowman2013-ka}%
  \BibitemOpen
  \bibfield  {author} {\bibinfo {author} {\bibfnamefont {G.~R.}\ \bibnamefont {Bowman}}, \bibinfo {author} {\bibfnamefont {V.~S.}\ \bibnamefont {Pande}}, \ and\ \bibinfo {author} {\bibfnamefont {F.}~\bibnamefont {No{\'e}}},\ }\href@noop {} {\emph {\bibinfo {title} {An introduction to Markov state models and their application to long timescale molecular simulation}}},\ Vol.\ \bibinfo {volume} {797}\ (\bibinfo  {publisher} {Springer Science \& Business Media},\ \bibinfo {year} {2013})\BibitemShut {NoStop}%
\bibitem [{\citenamefont {Husic}\ and\ \citenamefont {Pande}(2018)}]{husic2018markov}%
  \BibitemOpen
  \bibfield  {author} {\bibinfo {author} {\bibfnamefont {B.~E.}\ \bibnamefont {Husic}}\ and\ \bibinfo {author} {\bibfnamefont {V.~S.}\ \bibnamefont {Pande}},\ }\bibfield  {title} {\enquote {\bibinfo {title} {Markov state models: From an art to a science},}\ }\href@noop {} {\bibfield  {journal} {\bibinfo  {journal} {Journal of the American Chemical Society}\ }\textbf {\bibinfo {volume} {140}},\ \bibinfo {pages} {2386--2396} (\bibinfo {year} {2018})}\BibitemShut {NoStop}%
\bibitem [{\citenamefont {Nelson}\ and\ \citenamefont {Nelson}(2004)}]{nelson2004biological}%
  \BibitemOpen
  \bibfield  {author} {\bibinfo {author} {\bibfnamefont {P.~C.}\ \bibnamefont {Nelson}}\ and\ \bibinfo {author} {\bibfnamefont {P.}~\bibnamefont {Nelson}},\ }\href@noop {} {\emph {\bibinfo {title} {Biological physics}}}\ (\bibinfo  {publisher} {WH Freeman New York},\ \bibinfo {year} {2004})\BibitemShut {NoStop}%
\bibitem [{\citenamefont {Alon}(2019)}]{alon2019introduction}%
  \BibitemOpen
  \bibfield  {author} {\bibinfo {author} {\bibfnamefont {U.}~\bibnamefont {Alon}},\ }\href@noop {} {\emph {\bibinfo {title} {An introduction to systems biology: design principles of biological circuits}}}\ (\bibinfo  {publisher} {Chapman and Hall/CRC},\ \bibinfo {year} {2019})\BibitemShut {NoStop}%
\bibitem [{\citenamefont {Gillespie}(2001)}]{gillespie2001approximate}%
  \BibitemOpen
  \bibfield  {author} {\bibinfo {author} {\bibfnamefont {D.~T.}\ \bibnamefont {Gillespie}},\ }\bibfield  {title} {\enquote {\bibinfo {title} {Approximate accelerated stochastic simulation of chemically reacting systems},}\ }\href@noop {} {\bibfield  {journal} {\bibinfo  {journal} {The Journal of chemical physics}\ }\textbf {\bibinfo {volume} {115}},\ \bibinfo {pages} {1716--1733} (\bibinfo {year} {2001})}\BibitemShut {NoStop}%
\bibitem [{\citenamefont {Iglesias}\ and\ \citenamefont {Ingalls}(2010)}]{iglesias2010control}%
  \BibitemOpen
  \bibfield  {author} {\bibinfo {author} {\bibfnamefont {P.~A.}\ \bibnamefont {Iglesias}}\ and\ \bibinfo {author} {\bibfnamefont {B.~P.}\ \bibnamefont {Ingalls}},\ }\href@noop {} {\emph {\bibinfo {title} {Control theory and systems biology}}}\ (\bibinfo  {publisher} {MIT press},\ \bibinfo {year} {2010})\BibitemShut {NoStop}%
\bibitem [{\citenamefont {Lorendeau}\ \emph {et~al.}(2015)\citenamefont {Lorendeau}, \citenamefont {Christen}, \citenamefont {Rinaldi},\ and\ \citenamefont {Fendt}}]{lorendeau2015metabolic}%
  \BibitemOpen
  \bibfield  {author} {\bibinfo {author} {\bibfnamefont {D.}~\bibnamefont {Lorendeau}}, \bibinfo {author} {\bibfnamefont {S.}~\bibnamefont {Christen}}, \bibinfo {author} {\bibfnamefont {G.}~\bibnamefont {Rinaldi}}, \ and\ \bibinfo {author} {\bibfnamefont {S.-M.}\ \bibnamefont {Fendt}},\ }\bibfield  {title} {\enquote {\bibinfo {title} {Metabolic control of signalling pathways and metabolic auto-regulation},}\ }\href@noop {} {\bibfield  {journal} {\bibinfo  {journal} {Biology of the Cell}\ }\textbf {\bibinfo {volume} {107}},\ \bibinfo {pages} {251--272} (\bibinfo {year} {2015})}\BibitemShut {NoStop}%
\bibitem [{\citenamefont {Atkinson}(1965)}]{atkinson1965biological}%
  \BibitemOpen
  \bibfield  {author} {\bibinfo {author} {\bibfnamefont {D.~E.}\ \bibnamefont {Atkinson}},\ }\bibfield  {title} {\enquote {\bibinfo {title} {Biological feedback control at the molecular level: Interaction between metabolite-modulated enzymes seems to be a major factor in metabolic regulation.}}\ }\href@noop {} {\bibfield  {journal} {\bibinfo  {journal} {Science}\ }\textbf {\bibinfo {volume} {150}},\ \bibinfo {pages} {851--857} (\bibinfo {year} {1965})}\BibitemShut {NoStop}%
\bibitem [{\citenamefont {Goyal}\ \emph {et~al.}(2010)\citenamefont {Goyal}, \citenamefont {Yuan}, \citenamefont {Chen}, \citenamefont {Rabinowitz},\ and\ \citenamefont {Wingreen}}]{Goyal2010-ru}%
  \BibitemOpen
  \bibfield  {author} {\bibinfo {author} {\bibfnamefont {S.}~\bibnamefont {Goyal}}, \bibinfo {author} {\bibfnamefont {J.}~\bibnamefont {Yuan}}, \bibinfo {author} {\bibfnamefont {T.}~\bibnamefont {Chen}}, \bibinfo {author} {\bibfnamefont {J.~D.}\ \bibnamefont {Rabinowitz}}, \ and\ \bibinfo {author} {\bibfnamefont {N.~S.}\ \bibnamefont {Wingreen}},\ }\bibfield  {title} {\enquote {\bibinfo {title} {Achieving optimal growth through product feedback inhibition in metabolism},}\ }\href@noop {} {\bibfield  {journal} {\bibinfo  {journal} {PLoS Comput. Biol.}\ }\textbf {\bibinfo {volume} {6}},\ \bibinfo {pages} {e1000802} (\bibinfo {year} {2010})}\BibitemShut {NoStop}%
\bibitem [{\citenamefont {Yi}\ \emph {et~al.}(2000{\natexlab{a}})\citenamefont {Yi}, \citenamefont {Huang}, \citenamefont {Simon},\ and\ \citenamefont {Doyle}}]{yi2000robust}%
  \BibitemOpen
  \bibfield  {author} {\bibinfo {author} {\bibfnamefont {T.-M.}\ \bibnamefont {Yi}}, \bibinfo {author} {\bibfnamefont {Y.}~\bibnamefont {Huang}}, \bibinfo {author} {\bibfnamefont {M.~I.}\ \bibnamefont {Simon}}, \ and\ \bibinfo {author} {\bibfnamefont {J.}~\bibnamefont {Doyle}},\ }\bibfield  {title} {\enquote {\bibinfo {title} {Robust perfect adaptation in bacterial chemotaxis through integral feedback control},}\ }\href@noop {} {\bibfield  {journal} {\bibinfo  {journal} {Proceedings of the National Academy of Sciences}\ }\textbf {\bibinfo {volume} {97}},\ \bibinfo {pages} {4649--4653} (\bibinfo {year} {2000}{\natexlab{a}})}\BibitemShut {NoStop}%
\bibitem [{\citenamefont {Hamadeh}\ \emph {et~al.}(2011)\citenamefont {Hamadeh}, \citenamefont {Roberts}, \citenamefont {August}, \citenamefont {McSharry}, \citenamefont {Maini}, \citenamefont {Armitage},\ and\ \citenamefont {Papachristodoulou}}]{Hamadeh2011-hr}%
  \BibitemOpen
  \bibfield  {author} {\bibinfo {author} {\bibfnamefont {A.}~\bibnamefont {Hamadeh}}, \bibinfo {author} {\bibfnamefont {M.~A.~J.}\ \bibnamefont {Roberts}}, \bibinfo {author} {\bibfnamefont {E.}~\bibnamefont {August}}, \bibinfo {author} {\bibfnamefont {P.~E.}\ \bibnamefont {McSharry}}, \bibinfo {author} {\bibfnamefont {P.~K.}\ \bibnamefont {Maini}}, \bibinfo {author} {\bibfnamefont {J.~P.}\ \bibnamefont {Armitage}}, \ and\ \bibinfo {author} {\bibfnamefont {A.}~\bibnamefont {Papachristodoulou}},\ }\bibfield  {title} {\enquote {\bibinfo {title} {Feedback control architecture and the bacterial chemotaxis network},}\ }\href@noop {} {\bibfield  {journal} {\bibinfo  {journal} {PLoS Comput. Biol.}\ }\textbf {\bibinfo {volume} {7}},\ \bibinfo {pages} {e1001130} (\bibinfo {year} {2011})}\BibitemShut {NoStop}%
\bibitem [{\citenamefont {Yi}\ \emph {et~al.}(2000{\natexlab{b}})\citenamefont {Yi}, \citenamefont {Huang}, \citenamefont {Simon},\ and\ \citenamefont {Doyle}}]{Yi2000-pk}%
  \BibitemOpen
  \bibfield  {author} {\bibinfo {author} {\bibfnamefont {T.~M.}\ \bibnamefont {Yi}}, \bibinfo {author} {\bibfnamefont {Y.}~\bibnamefont {Huang}}, \bibinfo {author} {\bibfnamefont {M.~I.}\ \bibnamefont {Simon}}, \ and\ \bibinfo {author} {\bibfnamefont {J.}~\bibnamefont {Doyle}},\ }\bibfield  {title} {\enquote {\bibinfo {title} {Robust perfect adaptation in bacterial chemotaxis through integral feedback control},}\ }\href@noop {} {\bibfield  {journal} {\bibinfo  {journal} {Proc. Natl. Acad. Sci. U. S. A.}\ }\textbf {\bibinfo {volume} {97}},\ \bibinfo {pages} {4649--4653} (\bibinfo {year} {2000}{\natexlab{b}})}\BibitemShut {NoStop}%
\bibitem [{\citenamefont {Murray}(1992)}]{murray1992creative}%
  \BibitemOpen
  \bibfield  {author} {\bibinfo {author} {\bibfnamefont {A.~W.}\ \bibnamefont {Murray}},\ }\bibfield  {title} {\enquote {\bibinfo {title} {Creative blocks: cell-cycle checkpoints and feedback controls},}\ }\href@noop {} {\bibfield  {journal} {\bibinfo  {journal} {Nature}\ }\textbf {\bibinfo {volume} {359}},\ \bibinfo {pages} {599--601} (\bibinfo {year} {1992})}\BibitemShut {NoStop}%
\bibitem [{\citenamefont {Domian}, \citenamefont {Reisenauer},\ and\ \citenamefont {Shapiro}(1999)}]{domian1999feedback}%
  \BibitemOpen
  \bibfield  {author} {\bibinfo {author} {\bibfnamefont {I.~J.}\ \bibnamefont {Domian}}, \bibinfo {author} {\bibfnamefont {A.}~\bibnamefont {Reisenauer}}, \ and\ \bibinfo {author} {\bibfnamefont {L.}~\bibnamefont {Shapiro}},\ }\bibfield  {title} {\enquote {\bibinfo {title} {Feedback control of a master bacterial cell-cycle regulator},}\ }\href@noop {} {\bibfield  {journal} {\bibinfo  {journal} {Proceedings of the National Academy of Sciences}\ }\textbf {\bibinfo {volume} {96}},\ \bibinfo {pages} {6648--6653} (\bibinfo {year} {1999})}\BibitemShut {NoStop}%
\bibitem [{\citenamefont {Cross}\ and\ \citenamefont {Tinkelenberg}(1991)}]{cross1991potential}%
  \BibitemOpen
  \bibfield  {author} {\bibinfo {author} {\bibfnamefont {F.~R.}\ \bibnamefont {Cross}}\ and\ \bibinfo {author} {\bibfnamefont {A.~H.}\ \bibnamefont {Tinkelenberg}},\ }\bibfield  {title} {\enquote {\bibinfo {title} {A potential positive feedback loop controlling cln1 and cln2 gene expression at the start of the yeast cell cycle},}\ }\href@noop {} {\bibfield  {journal} {\bibinfo  {journal} {Cell}\ }\textbf {\bibinfo {volume} {65}},\ \bibinfo {pages} {875--883} (\bibinfo {year} {1991})}\BibitemShut {NoStop}%
\bibitem [{\citenamefont {Li}\ and\ \citenamefont {Murray}(1991)}]{li1991feedback}%
  \BibitemOpen
  \bibfield  {author} {\bibinfo {author} {\bibfnamefont {R.}~\bibnamefont {Li}}\ and\ \bibinfo {author} {\bibfnamefont {A.~W.}\ \bibnamefont {Murray}},\ }\bibfield  {title} {\enquote {\bibinfo {title} {Feedback control of mitosis in budding yeast},}\ }\href@noop {} {\bibfield  {journal} {\bibinfo  {journal} {Cell}\ }\textbf {\bibinfo {volume} {66}},\ \bibinfo {pages} {519--531} (\bibinfo {year} {1991})}\BibitemShut {NoStop}%
\bibitem [{\citenamefont {Gillespie}(1977)}]{gillespie1977exact}%
  \BibitemOpen
  \bibfield  {author} {\bibinfo {author} {\bibfnamefont {D.~T.}\ \bibnamefont {Gillespie}},\ }\bibfield  {title} {\enquote {\bibinfo {title} {Exact stochastic simulation of coupled chemical reactions},}\ }\href@noop {} {\bibfield  {journal} {\bibinfo  {journal} {The journal of physical chemistry}\ }\textbf {\bibinfo {volume} {81}},\ \bibinfo {pages} {2340--2361} (\bibinfo {year} {1977})}\BibitemShut {NoStop}%
\bibitem [{\citenamefont {M{\"o}bus}\ and\ \citenamefont {Wolf}(2019)}]{Mobus2019-qx}%
  \BibitemOpen
  \bibfield  {author} {\bibinfo {author} {\bibfnamefont {T.}~\bibnamefont {M{\"o}bus}}\ and\ \bibinfo {author} {\bibfnamefont {M.~M.}\ \bibnamefont {Wolf}},\ }\bibfield  {title} {\enquote {\bibinfo {title} {Quantum zeno effect generalized},}\ }\href@noop {} {\bibfield  {journal} {\bibinfo  {journal} {J. Math. Phys.}\ }\textbf {\bibinfo {volume} {60}},\ \bibinfo {pages} {052201} (\bibinfo {year} {2019})}\BibitemShut {NoStop}%
\bibitem [{\citenamefont {Itano}\ \emph {et~al.}(1990)\citenamefont {Itano}, \citenamefont {Heinzen}, \citenamefont {Bollinger},\ and\ \citenamefont {Wineland}}]{itano1990quantum}%
  \BibitemOpen
  \bibfield  {author} {\bibinfo {author} {\bibfnamefont {W.~M.}\ \bibnamefont {Itano}}, \bibinfo {author} {\bibfnamefont {D.~J.}\ \bibnamefont {Heinzen}}, \bibinfo {author} {\bibfnamefont {J.~J.}\ \bibnamefont {Bollinger}}, \ and\ \bibinfo {author} {\bibfnamefont {D.~J.}\ \bibnamefont {Wineland}},\ }\bibfield  {title} {\enquote {\bibinfo {title} {Quantum zeno effect},}\ }\href@noop {} {\bibfield  {journal} {\bibinfo  {journal} {Physical Review A}\ }\textbf {\bibinfo {volume} {41}},\ \bibinfo {pages} {2295} (\bibinfo {year} {1990})}\BibitemShut {NoStop}%
\bibitem [{\citenamefont {Streed}\ \emph {et~al.}(2006)\citenamefont {Streed}, \citenamefont {Mun}, \citenamefont {Boyd}, \citenamefont {Campbell}, \citenamefont {Medley}, \citenamefont {Ketterle},\ and\ \citenamefont {Pritchard}}]{streed2006continuous}%
  \BibitemOpen
  \bibfield  {author} {\bibinfo {author} {\bibfnamefont {E.~W.}\ \bibnamefont {Streed}}, \bibinfo {author} {\bibfnamefont {J.}~\bibnamefont {Mun}}, \bibinfo {author} {\bibfnamefont {M.}~\bibnamefont {Boyd}}, \bibinfo {author} {\bibfnamefont {G.~K.}\ \bibnamefont {Campbell}}, \bibinfo {author} {\bibfnamefont {P.}~\bibnamefont {Medley}}, \bibinfo {author} {\bibfnamefont {W.}~\bibnamefont {Ketterle}}, \ and\ \bibinfo {author} {\bibfnamefont {D.~E.}\ \bibnamefont {Pritchard}},\ }\bibfield  {title} {\enquote {\bibinfo {title} {Continuous and pulsed quantum zeno effect},}\ }\href@noop {} {\bibfield  {journal} {\bibinfo  {journal} {Physical review letters}\ }\textbf {\bibinfo {volume} {97}},\ \bibinfo {pages} {260402} (\bibinfo {year} {2006})}\BibitemShut {NoStop}%
\bibitem [{\citenamefont {Anderson}, \citenamefont {Ermentrout},\ and\ \citenamefont {Thomas}(2015)}]{Anderson2015-ql}%
  \BibitemOpen
  \bibfield  {author} {\bibinfo {author} {\bibfnamefont {D.~F.}\ \bibnamefont {Anderson}}, \bibinfo {author} {\bibfnamefont {B.}~\bibnamefont {Ermentrout}}, \ and\ \bibinfo {author} {\bibfnamefont {P.~J.}\ \bibnamefont {Thomas}},\ }\bibfield  {title} {\enquote {\bibinfo {title} {Stochastic representations of ion channel kinetics and exact stochastic simulation of neuronal dynamics},}\ }\href@noop {} {\bibfield  {journal} {\bibinfo  {journal} {Journal of computational neuroscience}\ }\textbf {\bibinfo {volume} {38}},\ \bibinfo {pages} {67--82} (\bibinfo {year} {2015})}\BibitemShut {NoStop}%
\bibitem [{\citenamefont {Ding}\ \emph {et~al.}(2016)\citenamefont {Ding}, \citenamefont {Qian}, \citenamefont {Qian},\ and\ \citenamefont {Zhang}}]{Ding2016-qr}%
  \BibitemOpen
  \bibfield  {author} {\bibinfo {author} {\bibfnamefont {S.}~\bibnamefont {Ding}}, \bibinfo {author} {\bibfnamefont {M.}~\bibnamefont {Qian}}, \bibinfo {author} {\bibfnamefont {H.}~\bibnamefont {Qian}}, \ and\ \bibinfo {author} {\bibfnamefont {X.}~\bibnamefont {Zhang}},\ }\bibfield  {title} {\enquote {\bibinfo {title} {Numerical simulations of piecewise deterministic markov processes with an application to the stochastic {Hodgkin-Huxley} model},}\ }\href@noop {} {\bibfield  {journal} {\bibinfo  {journal} {J. Chem. Phys.}\ }\textbf {\bibinfo {volume} {145}},\ \bibinfo {pages} {244107} (\bibinfo {year} {2016})}\BibitemShut {NoStop}%
\bibitem [{\citenamefont {Chow}\ and\ \citenamefont {White}(1996)}]{Chow1996-ab}%
  \BibitemOpen
  \bibfield  {author} {\bibinfo {author} {\bibfnamefont {C.~C.}\ \bibnamefont {Chow}}\ and\ \bibinfo {author} {\bibfnamefont {J.~A.}\ \bibnamefont {White}},\ }\bibfield  {title} {\enquote {\bibinfo {title} {Spontaneous action potentials due to channel fluctuations},}\ }\href@noop {} {\bibfield  {journal} {\bibinfo  {journal} {Biophys. J.}\ }\textbf {\bibinfo {volume} {71}},\ \bibinfo {pages} {3013--3021} (\bibinfo {year} {1996})}\BibitemShut {NoStop}%
\bibitem [{\citenamefont {Li}\ \emph {et~al.}(2009)\citenamefont {Li}, \citenamefont {Qu}, \citenamefont {Ma}, \citenamefont {Smith}, \citenamefont {Scherer},\ and\ \citenamefont {Dinner}}]{Li2009-ee}%
  \BibitemOpen
  \bibfield  {author} {\bibinfo {author} {\bibfnamefont {Y.}~\bibnamefont {Li}}, \bibinfo {author} {\bibfnamefont {X.}~\bibnamefont {Qu}}, \bibinfo {author} {\bibfnamefont {A.}~\bibnamefont {Ma}}, \bibinfo {author} {\bibfnamefont {G.~J.}\ \bibnamefont {Smith}}, \bibinfo {author} {\bibfnamefont {N.~F.}\ \bibnamefont {Scherer}}, \ and\ \bibinfo {author} {\bibfnamefont {A.~R.}\ \bibnamefont {Dinner}},\ }\bibfield  {title} {\enquote {\bibinfo {title} {Models of single-molecule experiments with periodic perturbations reveal hidden dynamics in rna folding},}\ }\href@noop {} {\bibfield  {journal} {\bibinfo  {journal} {The Journal of Physical Chemistry B}\ }\textbf {\bibinfo {volume} {113}},\ \bibinfo {pages} {7579--7590} (\bibinfo {year} {2009})}\BibitemShut {NoStop}%
\bibitem [{\citenamefont {Anderson}(2007)}]{Anderson2007-dq}%
  \BibitemOpen
  \bibfield  {author} {\bibinfo {author} {\bibfnamefont {D.~F.}\ \bibnamefont {Anderson}},\ }\bibfield  {title} {\enquote {\bibinfo {title} {A modified next reaction method for simulating chemical systems with time dependent propensities and delays},}\ }\href@noop {} {\bibfield  {journal} {\bibinfo  {journal} {J. Chem. Phys.}\ }\textbf {\bibinfo {volume} {127}},\ \bibinfo {pages} {214107} (\bibinfo {year} {2007})}\BibitemShut {NoStop}%
\bibitem [{\citenamefont {Bechhoefer}(2021)}]{bechhoefer2021control}%
  \BibitemOpen
  \bibfield  {author} {\bibinfo {author} {\bibfnamefont {J.}~\bibnamefont {Bechhoefer}},\ }\href@noop {} {\emph {\bibinfo {title} {Control theory for physicists}}}\ (\bibinfo  {publisher} {Cambridge University Press},\ \bibinfo {year} {2021})\BibitemShut {NoStop}%
\bibitem [{\citenamefont {Franklin}\ \emph {et~al.}(2002)\citenamefont {Franklin}, \citenamefont {Powell}, \citenamefont {Emami-Naeini},\ and\ \citenamefont {Powell}}]{franklin2002feedback}%
  \BibitemOpen
  \bibfield  {author} {\bibinfo {author} {\bibfnamefont {G.~F.}\ \bibnamefont {Franklin}}, \bibinfo {author} {\bibfnamefont {J.~D.}\ \bibnamefont {Powell}}, \bibinfo {author} {\bibfnamefont {A.}~\bibnamefont {Emami-Naeini}}, \ and\ \bibinfo {author} {\bibfnamefont {J.~D.}\ \bibnamefont {Powell}},\ }\href@noop {} {\emph {\bibinfo {title} {Feedback control of dynamic systems}}},\ Vol.~\bibinfo {volume} {4}\ (\bibinfo  {publisher} {Prentice hall Upper Saddle River},\ \bibinfo {year} {2002})\BibitemShut {NoStop}%
\bibitem [{Note1()}]{Note1}%
  \BibitemOpen
  \bibinfo {note} {Here constant rate over time is used to describe the time period between two consecutive stochastic events. Naturally, when the system state changes after the next event, the possible transitions and their rates are naturally updated.}\BibitemShut {Stop}%
\bibitem [{Note2()}]{Note2}%
  \BibitemOpen
  \bibinfo {note} {In practical implementation of the numerical code, when $q_i \approx q_{i-1}$, the quadratic equation reduces to the linear, which can be solved from \protect \cref {eqn:piecewise t*}.}\BibitemShut {Stop}%
\bibitem [{Note3()}]{Note3}%
  \BibitemOpen
  \bibinfo {note} {Here the measurement of the system may not be as detailed as the specific state that the system is in. It is possible that a subset of states will give a same measurement outcome (i.e., coarse-grained measurement). The method can be applied to both cases.}\BibitemShut {Stop}%
\bibitem [{\citenamefont {Annby-Andersson}\ \emph {et~al.}(2022)\citenamefont {Annby-Andersson}, \citenamefont {Bakhshinezhad}, \citenamefont {Bhattacharyya}, \citenamefont {De~Sousa}, \citenamefont {Jarzynski}, \citenamefont {Samuelsson},\ and\ \citenamefont {Potts}}]{annby2022quantum}%
  \BibitemOpen
  \bibfield  {author} {\bibinfo {author} {\bibfnamefont {B.}~\bibnamefont {Annby-Andersson}}, \bibinfo {author} {\bibfnamefont {F.}~\bibnamefont {Bakhshinezhad}}, \bibinfo {author} {\bibfnamefont {D.}~\bibnamefont {Bhattacharyya}}, \bibinfo {author} {\bibfnamefont {G.}~\bibnamefont {De~Sousa}}, \bibinfo {author} {\bibfnamefont {C.}~\bibnamefont {Jarzynski}}, \bibinfo {author} {\bibfnamefont {P.}~\bibnamefont {Samuelsson}}, \ and\ \bibinfo {author} {\bibfnamefont {P.~P.}\ \bibnamefont {Potts}},\ }\bibfield  {title} {\enquote {\bibinfo {title} {Quantum fokker-planck master equation for continuous feedback control},}\ }\href@noop {} {\bibfield  {journal} {\bibinfo  {journal} {Physical Review Letters}\ }\textbf {\bibinfo {volume} {129}},\ \bibinfo {pages} {050401} (\bibinfo {year} {2022})}\BibitemShut {NoStop}%
\bibitem [{\citenamefont {Horowitz}\ and\ \citenamefont {Jacobs}(2014)}]{horowitz2014quantum}%
  \BibitemOpen
  \bibfield  {author} {\bibinfo {author} {\bibfnamefont {J.~M.}\ \bibnamefont {Horowitz}}\ and\ \bibinfo {author} {\bibfnamefont {K.}~\bibnamefont {Jacobs}},\ }\bibfield  {title} {\enquote {\bibinfo {title} {Quantum effects improve the energy efficiency of feedback control},}\ }\href@noop {} {\bibfield  {journal} {\bibinfo  {journal} {Physical Review E}\ }\textbf {\bibinfo {volume} {89}},\ \bibinfo {pages} {042134} (\bibinfo {year} {2014})}\BibitemShut {NoStop}%
\bibitem [{\citenamefont {Horowitz}\ and\ \citenamefont {Parrondo}(2011{\natexlab{a}})}]{horowitz2011thermodynamic}%
  \BibitemOpen
  \bibfield  {author} {\bibinfo {author} {\bibfnamefont {J.~M.}\ \bibnamefont {Horowitz}}\ and\ \bibinfo {author} {\bibfnamefont {J.~M.}\ \bibnamefont {Parrondo}},\ }\bibfield  {title} {\enquote {\bibinfo {title} {Thermodynamic reversibility in feedback processes},}\ }\href@noop {} {\bibfield  {journal} {\bibinfo  {journal} {Europhysics Letters}\ }\textbf {\bibinfo {volume} {95}},\ \bibinfo {pages} {10005} (\bibinfo {year} {2011}{\natexlab{a}})}\BibitemShut {NoStop}%
\bibitem [{\citenamefont {Horowitz}\ and\ \citenamefont {Parrondo}(2011{\natexlab{b}})}]{horowitz2011designing}%
  \BibitemOpen
  \bibfield  {author} {\bibinfo {author} {\bibfnamefont {J.~M.}\ \bibnamefont {Horowitz}}\ and\ \bibinfo {author} {\bibfnamefont {J.~M.}\ \bibnamefont {Parrondo}},\ }\bibfield  {title} {\enquote {\bibinfo {title} {Designing optimal discrete-feedback thermodynamic engines},}\ }\href@noop {} {\bibfield  {journal} {\bibinfo  {journal} {New Journal of Physics}\ }\textbf {\bibinfo {volume} {13}},\ \bibinfo {pages} {123019} (\bibinfo {year} {2011}{\natexlab{b}})}\BibitemShut {NoStop}%
\bibitem [{\citenamefont {Horowitz}\ and\ \citenamefont {Vaikuntanathan}(2010)}]{horowitz2010nonequilibrium}%
  \BibitemOpen
  \bibfield  {author} {\bibinfo {author} {\bibfnamefont {J.~M.}\ \bibnamefont {Horowitz}}\ and\ \bibinfo {author} {\bibfnamefont {S.}~\bibnamefont {Vaikuntanathan}},\ }\bibfield  {title} {\enquote {\bibinfo {title} {Nonequilibrium detailed fluctuation theorem for repeated discrete feedback},}\ }\href@noop {} {\bibfield  {journal} {\bibinfo  {journal} {Physical Review E}\ }\textbf {\bibinfo {volume} {82}},\ \bibinfo {pages} {061120} (\bibinfo {year} {2010})}\BibitemShut {NoStop}%
\bibitem [{\citenamefont {Bhattacharyya}\ and\ \citenamefont {Jarzynski}(2022)}]{bhattacharyya2022feedback}%
  \BibitemOpen
  \bibfield  {author} {\bibinfo {author} {\bibfnamefont {D.}~\bibnamefont {Bhattacharyya}}\ and\ \bibinfo {author} {\bibfnamefont {C.}~\bibnamefont {Jarzynski}},\ }\bibfield  {title} {\enquote {\bibinfo {title} {From a feedback-controlled demon to an information ratchet in a double quantum dot},}\ }\href@noop {} {\bibfield  {journal} {\bibinfo  {journal} {Physical Review E}\ }\textbf {\bibinfo {volume} {106}},\ \bibinfo {pages} {064101} (\bibinfo {year} {2022})}\BibitemShut {NoStop}%
\bibitem [{\citenamefont {Sagawa}\ and\ \citenamefont {Ueda}(2012{\natexlab{a}})}]{sagawa2012nonequilibrium}%
  \BibitemOpen
  \bibfield  {author} {\bibinfo {author} {\bibfnamefont {T.}~\bibnamefont {Sagawa}}\ and\ \bibinfo {author} {\bibfnamefont {M.}~\bibnamefont {Ueda}},\ }\bibfield  {title} {\enquote {\bibinfo {title} {Nonequilibrium thermodynamics of feedback control},}\ }\href@noop {} {\bibfield  {journal} {\bibinfo  {journal} {Physical Review E}\ }\textbf {\bibinfo {volume} {85}},\ \bibinfo {pages} {021104} (\bibinfo {year} {2012}{\natexlab{a}})}\BibitemShut {NoStop}%
\bibitem [{\citenamefont {Sagawa}\ and\ \citenamefont {Ueda}(2010)}]{Sagawa2010-zv}%
  \BibitemOpen
  \bibfield  {author} {\bibinfo {author} {\bibfnamefont {T.}~\bibnamefont {Sagawa}}\ and\ \bibinfo {author} {\bibfnamefont {M.}~\bibnamefont {Ueda}},\ }\bibfield  {title} {\enquote {\bibinfo {title} {Generalized jarzynski equality under nonequilibrium feedback control},}\ }\href@noop {} {\bibfield  {journal} {\bibinfo  {journal} {Physical review letters}\ }\textbf {\bibinfo {volume} {104}},\ \bibinfo {pages} {090602} (\bibinfo {year} {2010})}\BibitemShut {NoStop}%
\bibitem [{\citenamefont {Sagawa}\ and\ \citenamefont {Ueda}(2012{\natexlab{b}})}]{sagawa2012fluctuation}%
  \BibitemOpen
  \bibfield  {author} {\bibinfo {author} {\bibfnamefont {T.}~\bibnamefont {Sagawa}}\ and\ \bibinfo {author} {\bibfnamefont {M.}~\bibnamefont {Ueda}},\ }\bibfield  {title} {\enquote {\bibinfo {title} {Fluctuation theorem with information exchange: Role of correlations in stochastic thermodynamics},}\ }\href@noop {} {\bibfield  {journal} {\bibinfo  {journal} {Physical review letters}\ }\textbf {\bibinfo {volume} {109}},\ \bibinfo {pages} {180602} (\bibinfo {year} {2012}{\natexlab{b}})}\BibitemShut {NoStop}%
\bibitem [{\citenamefont {Parrondo}, \citenamefont {Horowitz},\ and\ \citenamefont {Sagawa}(2015)}]{parrondo2015thermodynamics}%
  \BibitemOpen
  \bibfield  {author} {\bibinfo {author} {\bibfnamefont {J.~M.}\ \bibnamefont {Parrondo}}, \bibinfo {author} {\bibfnamefont {J.~M.}\ \bibnamefont {Horowitz}}, \ and\ \bibinfo {author} {\bibfnamefont {T.}~\bibnamefont {Sagawa}},\ }\bibfield  {title} {\enquote {\bibinfo {title} {Thermodynamics of information},}\ }\href@noop {} {\bibfield  {journal} {\bibinfo  {journal} {Nature physics}\ }\textbf {\bibinfo {volume} {11}},\ \bibinfo {pages} {131--139} (\bibinfo {year} {2015})}\BibitemShut {NoStop}%
\bibitem [{\citenamefont {Mandal}, \citenamefont {Quan},\ and\ \citenamefont {Jarzynski}(2013)}]{mandal2013maxwell}%
  \BibitemOpen
  \bibfield  {author} {\bibinfo {author} {\bibfnamefont {D.}~\bibnamefont {Mandal}}, \bibinfo {author} {\bibfnamefont {H.}~\bibnamefont {Quan}}, \ and\ \bibinfo {author} {\bibfnamefont {C.}~\bibnamefont {Jarzynski}},\ }\bibfield  {title} {\enquote {\bibinfo {title} {Maxwell’s refrigerator: an exactly solvable model},}\ }\href@noop {} {\bibfield  {journal} {\bibinfo  {journal} {Physical review letters}\ }\textbf {\bibinfo {volume} {111}},\ \bibinfo {pages} {030602} (\bibinfo {year} {2013})}\BibitemShut {NoStop}%
\bibitem [{\citenamefont {Sekimoto}()}]{Sekimoto_undated-xn}%
  \BibitemOpen
  \bibfield  {author} {\bibinfo {author} {\bibfnamefont {K.}~\bibnamefont {Sekimoto}},\ }\href@noop {} {\emph {\bibinfo {title} {Stochastic Energetics}}}\ (\bibinfo  {publisher} {Springer Berlin Heidelberg})\BibitemShut {NoStop}%
\bibitem [{\citenamefont {Penrose}(2005)}]{penrose2005foundations}%
  \BibitemOpen
  \bibfield  {author} {\bibinfo {author} {\bibfnamefont {O.}~\bibnamefont {Penrose}},\ }\href@noop {} {\emph {\bibinfo {title} {Foundations of statistical mechanics: a deductive treatment}}}\ (\bibinfo  {publisher} {Courier Corporation},\ \bibinfo {year} {2005})\BibitemShut {NoStop}%
\bibitem [{\citenamefont {Bennett}(1973)}]{bennett1973logical}%
  \BibitemOpen
  \bibfield  {author} {\bibinfo {author} {\bibfnamefont {C.~H.}\ \bibnamefont {Bennett}},\ }\bibfield  {title} {\enquote {\bibinfo {title} {Logical reversibility of computation},}\ }\href@noop {} {\bibfield  {journal} {\bibinfo  {journal} {IBM journal of Research and Development}\ }\textbf {\bibinfo {volume} {17}},\ \bibinfo {pages} {525--532} (\bibinfo {year} {1973})}\BibitemShut {NoStop}%
\bibitem [{\citenamefont {Landauer}(1961)}]{landauer1961irreversibility}%
  \BibitemOpen
  \bibfield  {author} {\bibinfo {author} {\bibfnamefont {R.}~\bibnamefont {Landauer}},\ }\bibfield  {title} {\enquote {\bibinfo {title} {Irreversibility and heat generation in the computing process},}\ }\href@noop {} {\bibfield  {journal} {\bibinfo  {journal} {IBM journal of research and development}\ }\textbf {\bibinfo {volume} {5}},\ \bibinfo {pages} {183--191} (\bibinfo {year} {1961})}\BibitemShut {NoStop}%
\bibitem [{\citenamefont {Mandal}\ and\ \citenamefont {Jarzynski}(2012)}]{mandal2012work}%
  \BibitemOpen
  \bibfield  {author} {\bibinfo {author} {\bibfnamefont {D.}~\bibnamefont {Mandal}}\ and\ \bibinfo {author} {\bibfnamefont {C.}~\bibnamefont {Jarzynski}},\ }\bibfield  {title} {\enquote {\bibinfo {title} {Work and information processing in a solvable model of maxwell’s demon},}\ }\href@noop {} {\bibfield  {journal} {\bibinfo  {journal} {Proceedings of the National Academy of Sciences}\ }\textbf {\bibinfo {volume} {109}},\ \bibinfo {pages} {11641--11645} (\bibinfo {year} {2012})}\BibitemShut {NoStop}%
\bibitem [{\citenamefont {Zurek}(1989)}]{zurek1989thermodynamic}%
  \BibitemOpen
  \bibfield  {author} {\bibinfo {author} {\bibfnamefont {W.~H.}\ \bibnamefont {Zurek}},\ }\bibfield  {title} {\enquote {\bibinfo {title} {Thermodynamic cost of computation, algorithmic complexity and the information metric},}\ }\href@noop {} {\bibfield  {journal} {\bibinfo  {journal} {Nature}\ }\textbf {\bibinfo {volume} {341}},\ \bibinfo {pages} {119--124} (\bibinfo {year} {1989})}\BibitemShut {NoStop}%
\bibitem [{\citenamefont {Deffner}(2013)}]{deffner2013information}%
  \BibitemOpen
  \bibfield  {author} {\bibinfo {author} {\bibfnamefont {S.}~\bibnamefont {Deffner}},\ }\bibfield  {title} {\enquote {\bibinfo {title} {Information-driven current in a quantum maxwell demon},}\ }\href@noop {} {\bibfield  {journal} {\bibinfo  {journal} {Physical Review E}\ }\textbf {\bibinfo {volume} {88}},\ \bibinfo {pages} {062128} (\bibinfo {year} {2013})}\BibitemShut {NoStop}%
\bibitem [{\citenamefont {Strasberg}\ \emph {et~al.}(2013)\citenamefont {Strasberg}, \citenamefont {Schaller}, \citenamefont {Brandes},\ and\ \citenamefont {Esposito}}]{strasberg2013thermodynamics}%
  \BibitemOpen
  \bibfield  {author} {\bibinfo {author} {\bibfnamefont {P.}~\bibnamefont {Strasberg}}, \bibinfo {author} {\bibfnamefont {G.}~\bibnamefont {Schaller}}, \bibinfo {author} {\bibfnamefont {T.}~\bibnamefont {Brandes}}, \ and\ \bibinfo {author} {\bibfnamefont {M.}~\bibnamefont {Esposito}},\ }\bibfield  {title} {\enquote {\bibinfo {title} {Thermodynamics of a physical model implementing a maxwell demon},}\ }\href@noop {} {\bibfield  {journal} {\bibinfo  {journal} {Physical review letters}\ }\textbf {\bibinfo {volume} {110}},\ \bibinfo {pages} {040601} (\bibinfo {year} {2013})}\BibitemShut {NoStop}%
\bibitem [{\citenamefont {Lu}, \citenamefont {Mandal},\ and\ \citenamefont {Jarzynski}(2014)}]{lu2014engineering}%
  \BibitemOpen
  \bibfield  {author} {\bibinfo {author} {\bibfnamefont {Z.}~\bibnamefont {Lu}}, \bibinfo {author} {\bibfnamefont {D.}~\bibnamefont {Mandal}}, \ and\ \bibinfo {author} {\bibfnamefont {C.}~\bibnamefont {Jarzynski}},\ }\bibfield  {title} {\enquote {\bibinfo {title} {Engineering maxwell’s demon},}\ }\href@noop {} {\bibfield  {journal} {\bibinfo  {journal} {Physics Today}\ }\textbf {\bibinfo {volume} {67}},\ \bibinfo {pages} {60--61} (\bibinfo {year} {2014})}\BibitemShut {NoStop}%
\bibitem [{\citenamefont {Fischer}, \citenamefont {Guti{\'e}rrez-Medina},\ and\ \citenamefont {Raizen}(2001)}]{Fischer2001-bn}%
  \BibitemOpen
  \bibfield  {author} {\bibinfo {author} {\bibfnamefont {M.~C.}\ \bibnamefont {Fischer}}, \bibinfo {author} {\bibfnamefont {B.}~\bibnamefont {Guti{\'e}rrez-Medina}}, \ and\ \bibinfo {author} {\bibfnamefont {M.~G.}\ \bibnamefont {Raizen}},\ }\bibfield  {title} {\enquote {\bibinfo {title} {Observation of the quantum zeno and anti-zeno effects in an unstable system},}\ }\href@noop {} {\bibfield  {journal} {\bibinfo  {journal} {Phys. Rev. Lett.}\ }\textbf {\bibinfo {volume} {87}},\ \bibinfo {pages} {040402} (\bibinfo {year} {2001})}\BibitemShut {NoStop}%
\bibitem [{Note4()}]{Note4}%
  \BibitemOpen
  \bibinfo {note} {Under feedback control measurements, the system's kinetic rates can be updated after the measurement, but the system's state is not directly altered by the action of measurement.}\BibitemShut {Stop}%
\bibitem [{\citenamefont {Seidler}, \citenamefont {Noll},\ and\ \citenamefont {Thiers}(2004)}]{seidler2004feedforward}%
  \BibitemOpen
  \bibfield  {author} {\bibinfo {author} {\bibfnamefont {R.~D.}\ \bibnamefont {Seidler}}, \bibinfo {author} {\bibfnamefont {D.~C.}\ \bibnamefont {Noll}}, \ and\ \bibinfo {author} {\bibfnamefont {G.}~\bibnamefont {Thiers}},\ }\bibfield  {title} {\enquote {\bibinfo {title} {Feedforward and feedback processes in motor control},}\ }\href@noop {} {\bibfield  {journal} {\bibinfo  {journal} {Neuroimage}\ }\textbf {\bibinfo {volume} {22}},\ \bibinfo {pages} {1775--1783} (\bibinfo {year} {2004})}\BibitemShut {NoStop}%
\bibitem [{\citenamefont {Henninger}\ \emph {et~al.}(2021)\citenamefont {Henninger}, \citenamefont {Oksuz}, \citenamefont {Shrinivas}, \citenamefont {Sagi}, \citenamefont {LeRoy}, \citenamefont {Zheng}, \citenamefont {Andrews}, \citenamefont {Zamudio}, \citenamefont {Lazaris}, \citenamefont {Hannett} \emph {et~al.}}]{henninger2021rna}%
  \BibitemOpen
  \bibfield  {author} {\bibinfo {author} {\bibfnamefont {J.~E.}\ \bibnamefont {Henninger}}, \bibinfo {author} {\bibfnamefont {O.}~\bibnamefont {Oksuz}}, \bibinfo {author} {\bibfnamefont {K.}~\bibnamefont {Shrinivas}}, \bibinfo {author} {\bibfnamefont {I.}~\bibnamefont {Sagi}}, \bibinfo {author} {\bibfnamefont {G.}~\bibnamefont {LeRoy}}, \bibinfo {author} {\bibfnamefont {M.~M.}\ \bibnamefont {Zheng}}, \bibinfo {author} {\bibfnamefont {J.~O.}\ \bibnamefont {Andrews}}, \bibinfo {author} {\bibfnamefont {A.~V.}\ \bibnamefont {Zamudio}}, \bibinfo {author} {\bibfnamefont {C.}~\bibnamefont {Lazaris}}, \bibinfo {author} {\bibfnamefont {N.~M.}\ \bibnamefont {Hannett}},  \emph {et~al.},\ }\bibfield  {title} {\enquote {\bibinfo {title} {Rna-mediated feedback control of transcriptional condensates},}\ }\href@noop {} {\bibfield  {journal} {\bibinfo  {journal} {Cell}\ }\textbf {\bibinfo {volume} {184}},\ \bibinfo {pages} {207--225} (\bibinfo {year} {2021})}\BibitemShut {NoStop}%
\bibitem [{\citenamefont {Brown}\ and\ \citenamefont {Doyle~III}(2020)}]{brown2020dual}%
  \BibitemOpen
  \bibfield  {author} {\bibinfo {author} {\bibfnamefont {L.~S.}\ \bibnamefont {Brown}}\ and\ \bibinfo {author} {\bibfnamefont {F.~J.}\ \bibnamefont {Doyle~III}},\ }\bibfield  {title} {\enquote {\bibinfo {title} {A dual-feedback loop model of the mammalian circadian clock for multi-input control of circadian phase},}\ }\href@noop {} {\bibfield  {journal} {\bibinfo  {journal} {PLoS Computational Biology}\ }\textbf {\bibinfo {volume} {16}},\ \bibinfo {pages} {e1008459} (\bibinfo {year} {2020})}\BibitemShut {NoStop}%
\bibitem [{\citenamefont {Chaves}\ and\ \citenamefont {Oyarz{\'u}n}(2019)}]{chaves2019dynamics}%
  \BibitemOpen
  \bibfield  {author} {\bibinfo {author} {\bibfnamefont {M.}~\bibnamefont {Chaves}}\ and\ \bibinfo {author} {\bibfnamefont {D.~A.}\ \bibnamefont {Oyarz{\'u}n}},\ }\bibfield  {title} {\enquote {\bibinfo {title} {Dynamics of complex feedback architectures in metabolic pathways},}\ }\href@noop {} {\bibfield  {journal} {\bibinfo  {journal} {Automatica}\ }\textbf {\bibinfo {volume} {99}},\ \bibinfo {pages} {323--332} (\bibinfo {year} {2019})}\BibitemShut {NoStop}%
\bibitem [{\citenamefont {Tagliazucchi}, \citenamefont {Weiss},\ and\ \citenamefont {Szleifer}(2014)}]{tagliazucchi2014dissipative}%
  \BibitemOpen
  \bibfield  {author} {\bibinfo {author} {\bibfnamefont {M.}~\bibnamefont {Tagliazucchi}}, \bibinfo {author} {\bibfnamefont {E.~A.}\ \bibnamefont {Weiss}}, \ and\ \bibinfo {author} {\bibfnamefont {I.}~\bibnamefont {Szleifer}},\ }\bibfield  {title} {\enquote {\bibinfo {title} {Dissipative self-assembly of particles interacting through time-oscillatory potentials},}\ }\href@noop {} {\bibfield  {journal} {\bibinfo  {journal} {Proceedings of the National Academy of Sciences}\ }\textbf {\bibinfo {volume} {111}},\ \bibinfo {pages} {9751--9756} (\bibinfo {year} {2014})}\BibitemShut {NoStop}%
\bibitem [{\citenamefont {Zhang}, \citenamefont {Du},\ and\ \citenamefont {Lu}(2023)}]{zhang2023energy}%
  \BibitemOpen
  \bibfield  {author} {\bibinfo {author} {\bibfnamefont {Z.}~\bibnamefont {Zhang}}, \bibinfo {author} {\bibfnamefont {V.}~\bibnamefont {Du}}, \ and\ \bibinfo {author} {\bibfnamefont {Z.}~\bibnamefont {Lu}},\ }\bibfield  {title} {\enquote {\bibinfo {title} {Energy landscape design principle for optimal energy harnessing by catalytic molecular machines},}\ }\href@noop {} {\bibfield  {journal} {\bibinfo  {journal} {Physical Review E}\ }\textbf {\bibinfo {volume} {107}},\ \bibinfo {pages} {L012102} (\bibinfo {year} {2023})}\BibitemShut {NoStop}%
\end{thebibliography}%

\end{document}